\documentclass[11pt, reqno]{article}
\pdfoutput=1

\usepackage{jheppub}
\usepackage{amssymb}
\usepackage{amsmath}
\usepackage[usenames,dvipsnames]{xcolor}
\usepackage{epsfig}
\usepackage{dcolumn}
\usepackage{tikz}
\usetikzlibrary{shapes.geometric, arrows}
\usepackage{upgreek}
\usepackage{setspace}
\usepackage{enumitem}
\usepackage{array,multirow,bigdelim,arydshln}
\usepackage{appendix}
\usepackage{xparse}
\usepackage[bottom]{footmisc}
\usepackage[utf8]{inputenc}
\usepackage{microtype}
\hypersetup{
	colorlinks,
	urlcolor=Maroon,
	linkcolor=Maroon,
	citecolor=Maroon
}

\def \be {\begin{equation}}
\def \en {\end{equation}}
\def \bes {\begin{eqnarray}}
\def \ens {\end{eqnarray}}
\def \tr {\nonumber\\}

\def \nn {\nonumber}

\def \KLT {\,\overset{\mathrm{\scriptscriptstyle KLT}}{\otimes}\,}
\def \KLTs {\,\overset{\substack{\mathrm{\scriptscriptstyle string}\\ \mathrm{\scriptscriptstyle KLT}}}{\otimes}\,}

\title{Inverse of the String Theory KLT Kernel}
\author{Sebastian Mizera}
\affiliation{Perimeter Institute for Theoretical Physics, Waterloo, ON N2L 2Y5, Canada}
\affiliation{Department of Physics \& Astronomy, University of Waterloo, Waterloo, ON N2L 3G1, Canada}
\emailAdd{smizera@pitp.ca}
\abstract{The field theory Kawai--Lewellen--Tye (KLT) kernel, which relates scattering amplitudes of gravitons and gluons, turns out to be the inverse of a matrix whose components are bi-adjoint scalar partial amplitudes. In this note we propose an analogous construction for the string theory KLT kernel. We present simple diagrammatic rules for the computation of the $\alpha'$-corrected bi-adjoint scalar amplitudes that are exact in $\alpha'$. We find compact expressions in terms of graphs, where the standard Feynman propagators $1/p^2$ are replaced by either $1/\sin (\pi \alpha' p^2/2)$ or $1/\tan (\pi \alpha' p^2/2)$, as determined by a recursive procedure. We demonstrate how the same object can be used to conveniently expand open string partial amplitudes in a BCJ basis.}
\arxivnumber{1610.04230}

\begin{document}

\maketitle
\addtocontents{toc}{\protect\setcounter{tocdepth}{1}}

\section{Introduction}

In 1986, Kawai, Lewellen and Tye (KLT) revolutionized our understanding of the relations between scattering amplitudes \cite{Kawai:1985xq}. They discovered that tree-level amplitudes of closed strings can be written as a quadratic combination of the open string amplitudes. In the infinite tension limit, these relations tell us how to construct pure gravity amplitudes as a ``square'' of the Yang--Mills theory \cite{Bern:1998sv,BjerrumBohr:2010hn}.

Nearly 30 years later, Cachazo, He and Yuan (CHY) refined this statement by showing that the coefficients of the field-theory KLT expansion are given by the inverse of a matrix whose components are nothing but the bi-adjoint scalar amplitudes \cite{Cachazo:2013gna,Cachazo:2013hca,Cachazo:2013iea}. Schematically, we can write:\vspace{-1em}
\bes
\text{Gravity} = \frac{\text{Yang--Mills}^2}{\text{bi-adjoint scalar}}.\nn
\ens
We make this relationship precise in section \ref{sec:Review}, where we also give a brief review of the bi-adjoint scalar theory.

This new interpretation of the field-theory KLT kernel hints at a possible deeper understanding of the ``squaring'' procedure. A natural question is how it extends to the full string theory KLT kernel. In this note, we introduce a new object that generalizes the bi-adjoint scalar to include the $\alpha'$ dependence. We propose that it leads to the KLT relation:
\bes
\text{Closed string} = \frac{\text{Open string}^2}{\alpha'\text{-corrected bi-adjoint scalar}}.\nn
\ens
As it turns out, the new $\alpha'$-corrected amplitudes, which we denote by $m_{\alpha'}(\beta | \tilde{\beta})$, display an interesting structure. We will explore it in this work.

Before properly introducing the notation, let us take a glimpse on some examples of the amplitudes\footnote{From the form of (\ref{eq:m-alpha-1234-1234}--\ref{eq:m-alpha-12345-12345}) it is evident that the objects $m_{\alpha'}(\beta | \tilde{\beta})$ contain an infinite number of simple poles corresponding to massless, massive and tachyonic states. In a slight abuse of terminology, we will refer to them as {\emph{amplitudes}} of an $\alpha'$-corrected bi-adjoint scalar theory, since they compute physical quantities at the leading order in $\alpha'$. Nonetheless, we think of $m_{\alpha'}(\beta | \tilde{\beta})$ as a purely combinatorial object.} we will study,
\be\label{eq:m-alpha-1234-1234}
m_{\alpha'}(1234|1234) = \frac{1}{\tan (\pi \alpha' s_{12})} + \frac{1}{\tan (\pi \alpha' s_{23})},
\en
\be\label{eq:m-alpha-123456-126345}
m_{\alpha'}(123456|126345) = \frac{1}{\sin (\pi \alpha' s_{12}) \,\sin (\pi \alpha' s_{345})} \left(\frac{1}{\tan (\pi \alpha' s_{34})} + \frac{1}{\tan (\pi \alpha' s_{45})} \right),
\en
\bes\label{eq:m-alpha-12345-12345}
m_{\alpha'}(12345 | 12345 ) &=& \frac{1}{\tan (\pi \alpha' s_{12})\, \tan (\pi \alpha' s_{34})} + \frac{1}{\tan (\pi \alpha' s_{23})\, \tan (\pi \alpha' s_{45})}\tr
&& + \frac{1}{\tan (\pi \alpha' s_{34})\, \tan (\pi \alpha' s_{51})} + \frac{1}{\tan (\pi \alpha' s_{45})\, \tan (\pi \alpha' s_{12})} \tr
&&  + \frac{1}{\tan (\pi \alpha' s_{51})\, \tan (\pi \alpha' s_{23})} + 1.
\ens
Surprisingly, these objects have {\it compact} expressions in terms of trigonometric functions that are exact in $\alpha'$ and can be calculated using simple diagrammatic rules! We present these rules in section \ref{sec:Diagrammatic Rules}. We will see that they correctly recover the standard bi-adjoint theory in the infinite tension limit, $\alpha' \to 0$.

In section \ref{sec:KLT Relations}, we show how to construct string theory KLT relations based on the amplitudes in the $\alpha'$-corrected bi-adjoint theory. The KLT kernel will be given by the inverse of a matrix of the objects $m_{\alpha'}(\beta | \tilde{\beta})$. As the first application, we show how the newly-found interpretation of the KLT kernel can be used to derive the leading soft factors of closed string amplitudes from the open string ones.

In section \ref{sec:Basis Expansion}, we explain how the same structure can be used to generate a basis expansion of open string partial amplitudes. In the $\alpha' \to 0$ limit, the expansion gives the Bern--Carrasco--Johansson (BCJ) basis for the Yang--Mills partial amplitudes \cite{Bern:2008qj}.

It is important to mention that other string-like constructions that can be thought of as $\alpha'$-corrected versions of the bi-adjoint scalar have been previously studied in \cite{Mafra:2011nv,Mafra:2011nw,Broedel:2013aza,Broedel:2013tta,Stieberger:2014hba,Huang:2016tag,Carrasco:2016ldy,Mafra:2016mcc}. In particular, the objects $Z_{\beta}(\gamma)$ introduced in \cite{Mafra:2011nv,Mafra:2011nw} are used to construct open string amplitudes from super Yang--Mills amplitudes. In this case, the amplitudes $Z_{\beta}(\gamma)$ encapsulate all the $\alpha'$ dependence of the string scattering amplitudes. We connect it to the object $m_{\alpha'}(\beta | \tilde{\beta})$ studied in this work in section \ref{sec:Future Directions}, where we also discuss possible future directions.

As an ancillary file to this \texttt{arXiv} submission, we have attached a \texttt{Mathematica} notebook which allows to reproduce all $\alpha'$-corrected amplitudes $m_{\alpha'}(\beta | \tilde{\beta})$ studied in this work. It can be conveniently used to generate KLT and BCJ expansion coeffiecients in both string theory and field-theory.

\section{\label{sec:Review}Review of the Bi-adjoint Scalar Theory}

Before introducing the new theory, we review the field theory bi-adjoint scalar \cite{BjerrumBohr:2012mg}\cite{Cachazo:2013hca,Cachazo:2013iea}. It is a massless scalar field, $\phi_{a\tilde{a}}$, that derives its name from the two flavour groups, $U(N) \times U(\tilde{N})$, under which it transforms in the adjoint representation. The theory contains a single self-interaction of the form $f^{abc} \tilde{f}^{\tilde{a}\tilde{b}\tilde{c}}\, \phi^{}_{a\tilde{a}}\, \phi_{b\tilde{b}}\, \phi^{}_{c\tilde{c}}$, where $f^{abc}$ and $\tilde{f}^{\tilde{a}\tilde{b}\tilde{c}}$ are the structure constants of the two flavour groups.

The full scattering amplitude of $n$ bi-adjoint scalars can be expanded in the trace decomposition:
\be\label{eq:Trace decomposition}
m_{\text{full}} \;= \!\!\!\sum_{\beta, \tilde{\beta} \,\in\, S_n/\mathbb{Z}_n} \!\!\!\!\! \text{Tr}(T^{a_{\beta_1}} T^{a_{\beta_2}} \cdots T^{a_{\beta_n}})\, \text{Tr}(\tilde{T}^{\tilde{a}_{\tilde{\beta}_1}} \tilde{T}^{\tilde{a}_{\tilde{\beta}_2}} \cdots \tilde{T}^{\tilde{a}_{\tilde{\beta}_n}})\, m(\beta\, | \tilde{\beta}),
\en
where the sum goes over all the permutations $\beta$ and $\tilde{\beta}$ modulo cyclicity. The partial amplitudes, $m(\beta|\tilde{\beta})$, are the objects of our interest. They can be written as a sum over all trivalent graphs that are planar with respect to the first partial ordering, $\mathcal{G}_\beta$, and the second partial ordering, $\mathcal{G}_{\tilde{\beta}}$, at the same time. Every internal edge, $e$, is decorated with an appropriate propagator,
\be
m(\beta | \tilde{\beta})\; =\; \pm\!\!\!\!\sum_{g \,\in\, \mathcal{G}_\beta \,\cap\, \mathcal{G}_{\tilde{\beta}}} \frac{1}{\prod_{e \in g} s_e},
\en
where $s_e = p_e^2/2$ is the norm of the momentum flowing through the edge $e$. In particular, if there are no diagrams consistent with both permutations, the amplitude vanishes. The overall sign can be determined using the rules described in \cite{Cachazo:2013iea}. We will introduce an alternative construction later in section \ref{sec:Off-diagonal amplitudes}, after the discussion of the diagrammatic rules.

Examples of the amplitudes are as follows:
\be\label{eq:Example n=3}
m(123|123) = 1,\phantom{\frac{1}{1}}
\en
\be\label{eq:Example n=4}
m(1234|1234) = \frac{1}{s_{12}} + \frac{1}{s_{23}}, \qquad m(1234|1243) = -\frac{1}{s_{12}},
\en
\bes\label{eq:Example n=5}
m(12345|13254) = \frac{1}{s_{23} s_{45}}, &&\qquad m(12435|14253) = \frac{1}{s_{24}s_{35}},\tr\tr
m(12345|14253) = 0, &&\qquad m(12435|13254) = 0.\phantom{\frac{1}{1}}
\ens
In the last two cases there are no diagrams consistent with both orderings and hence the corresponding amplitudes vanish.

Using the CHY representation \cite{Cachazo:2013gna,Cachazo:2013hca,Cachazo:2013iea} of the bi-adjoint scalar, a simple linear algebra derivation \cite{Cachazo:2013iea} leads to the field theory KLT relation,
\bes\label{eq:KLT relation}
\mathcal{M}^{\text{GR}}_n &=& \mathcal{A}^{\text{YM}} \KLT \mathcal{A}^{\text{YM}}\tr
&\equiv& \mathcal{A}^{\text{YM}}(\beta) \; m^{-1}(\beta | \tilde{\beta})\; \mathcal{A}^{\text{YM}}(\tilde{\beta}),
\ens
which we take as a definition of the operation $\!\!\KLT\!\!$. Here, $\mathcal{M}^{\text{GR}}_n$ is a pure gravity amplitude and $\mathcal{A}(\beta)$ is a Yang--Mills partial amplitude obtained from the decomposition, $\mathcal{A}^{\text{YM}}_{\text{full}} = \sum_{\beta\in S_n/\mathbb{Z}_n} \text{Tr}(T^{a_{\beta_1}} T^{a_{\beta_2}} \cdots T^{a_{\beta_n}})\, \mathcal{A}^{\text{YM}}(\beta)$, analogous to (\ref{eq:Trace decomposition}). By $m^{-1}(\beta | \tilde{\beta})$ we mean the inverse of the matrix of bi-adjoint amplitudes, with columns and rows labelled by permutations $\beta$ and $\tilde{\beta}$ respectively. The vectors $\mathcal{A}^{\text{YM}}(\beta)$ and $\mathcal{A}^{\text{YM}}(\tilde{\beta})$ are defined similarly. The sum over repeated indices $\beta$ and $\tilde{\beta}$ is implied. Here $\beta$ and $\tilde{\beta}$ range over sets of $(n-3)!$ permutations forming a BCJ basis, in which case the $(n-3)! \times (n-3)!$ matrix $m^{-1}(\beta | \tilde{\beta})$ is invertible.

Let us illustrate this construction with an example. For $n=5$, we choose two sets of orderings to be $\beta \in \left\{ (12345), (12435)\right\}$ and $\tilde{\beta} \in \left\{ (13254), (14253)\right\}$. We can then use the amplitudes computed in (\ref{eq:Example n=5}) to construct the KLT kernel matrix as follows:
\bes\label{eq:KLT5}
\mathcal{M}_5^{\text{GR}} &=& \begin{bmatrix} \mathcal{A}^{\text{YM}}(12345)\\ \mathcal{A}^{\text{YM}}(12435) \end{bmatrix}^\intercal \begin{bmatrix} 1/s_{23}s_{45} & 0\\ 0 & 1/s_{24}s_{35} \end{bmatrix}^{-1} \begin{bmatrix} \mathcal{A}^{\text{YM}}(13254)\\ \mathcal{A}^{\text{YM}}(14253) \end{bmatrix}\tr
&=& s_{23} s_{45} \,\mathcal{A}^{\text{YM}}(12345)\, \mathcal{A}^{\text{YM}}(13254) + s_{24} s_{35}\, \mathcal{A}^{\text{YM}}(12435)\, \mathcal{A}^{\text{YM}}(14253).\phantom{\frac{1}{1}}
\ens
The bi-adjoint scalar can be used to build the matrix $m(\beta | \tilde{\beta})$ with arbitrary permutations, as long as the matrix is invertible.

Using the CHY representation, it was found that the KLT relation (\ref{eq:KLT relation}) generalizes to other theories as well \cite{Cachazo:2014nsa,Cachazo:2014xea,He:2016vfi,Cachazo:2016njl}. For instance, Born-Infeld theory can be written as a KLT product of Yang--Mills with the non-linear sigma model \cite{Cachazo:2014xea}. Despite the fact that the relations are linking different theories, the kernel stays the same.

In addition to this, the bi-adjoint scalar also generates a basis expansion of partial amplitudes \cite{Cachazo:2013iea}. Taking the example of the Yang--Mills theory, we can write:
\bes\label{eq:basis-expansion}
\mathcal{A}^{\text{YM}}(\beta) = m(\beta | \tilde{\beta})\, m^{-1}(\tilde{\beta} | \gamma)\, \mathcal{A}^{\text{YM}}(\gamma),
\ens
which in fact can be understood as a KLT of a bi-adjoint scalar with the Yang--Mills theory. Here, $\mathcal{A}^{\text{YM}}(\beta)$ on the left hand side is a vector of size $p$, $m(\beta | \tilde{\beta})$ is an $p \times (n-3)!$ matrix, $m^{-1}(\tilde{\beta} | \gamma)$ is an $(n-3)! \times (n-3)!$ matrix and finally $\mathcal{A}^{\text{YM}}(\gamma)$ is a vector of size $(n-3)!$. Hence, each of the $p$ partial amplitudes on the left hand side is being written as a linear combination of $(n-3)!$ Yang--Mills partial amplitudes. The latter forms a basis.

For instance, let us expand $\mathcal{A}^{\text{YM}}(12354)$ in the basis $\{ \mathcal{A}^{\text{YM}}(13254), \mathcal{A}^{\text{YM}}(14253)\}$. We can reuse the matrix from \eqref{eq:KLT5} to write:
\bes\label{eq:basis5}
\mathcal{A}^{\text{YM}}(12354) &=& \begin{bmatrix} - 1/s_{12}s_{45} - 1/s_{23}s_{45} \\ -1/s_{12} s_{35} \end{bmatrix}^\intercal \begin{bmatrix} 1/s_{23}s_{45} & 0\\ 0 & 1/s_{24}s_{35} \end{bmatrix}^{-1} \begin{bmatrix} \mathcal{A}^{\text{YM}}(13254)\\ \mathcal{A}^{\text{YM}}(14253) \end{bmatrix}\tr\tr
&=& - \frac{s_{12}+s_{23}}{s_{12}}\, \mathcal{A}^{\text{YM}}(13254) - \frac{s_{24}}{s_{12}}\, \mathcal{A}^{\text{YM}}(14253).
\ens
Here, we have also used two extra bi-adjoint amplitudes:
\be
m(12354|12345) = -\frac{1}{s_{12} s_{45}} - \frac{1}{s_{23}s_{45}}, \qquad m(12354|12435) = -\frac{1}{s_{12} s_{35}}.
\en
In similarity to the KLT relations, the basis expansion is valid in any dimension and for any choice of polarization vectors. The same expansion also extends to the supersymmetric case \cite{Stieberger:2009hq}.

The bi-adjoint scalar theory can be also used to understand the colour-kinematics duality \cite{Bern:2008qj,Bern:2010ue,Bern:2010yg,Elvang:2013cua} at tree level \cite{BjerrumBohr:2012mg,Cachazo:2013gna,Cachazo:2013iea,Monteiro:2013rya,Chiodaroli:2015rdg,Bjerrum-Bohr:2016axv}. Moreover, it was recently shown that this theory plays a crucial role in the understanding of the classical double copy relations  \cite{Monteiro:2014cda,Luna:2015paa,Luna:2016due,Ridgway:2015fdl} between exact solutions in Yang--Mills and gravity. The equations of motion have been studied in \cite{White:2016jzc}. Berends--Giele recursion relations were given in \cite{Mafra:2016ltu}.

\section{\label{sec:Diagrammatic Rules}Diagrammatic Rules for $m_{\alpha'}(\beta|\tilde{\beta})$}

In this section we investigate the analogue of (\ref{eq:KLT relation}) in string theory. Let us assume there exist some amplitudes that compute the inverse of the KLT kernel exactly in $\alpha'$, such that,\footnote{Following \cite{Cachazo:2013iea}, by $m_{\alpha'}^{-1}(\beta | \tilde{\beta})$ we denote the entries of a matrix $m_{\alpha'}(\tilde{\beta}|\beta)$ after taking the inverse.}
\bes\label{eq:string KLT}
\mathcal{M}^{\text{closed}}_n &=& \mathcal{A}^{\text{open}} \KLTs \mathcal{A}^{\text{open}}\tr
&\equiv& \mathcal{A}^{\text{open}}(\beta) \; m_{\alpha'}^{-1}(\beta | \tilde{\beta})\; \mathcal{A}^{\text{open}}(\tilde{\beta}).
\ens
Here $\!\!\KLTs\!\!$ is the original KLT operation \cite{Kawai:1985xq}, now defined in terms of a new object, $m_{\alpha'}(\beta | \tilde{\beta})$. Since the open and closed string amplitudes become the Yang--Mills and gravity ones in the $\alpha' \to 0$ limit, we require that the same happens for the $\alpha'$-corrected bi-adjoint theory. In our normalization,
\bes
m_{\alpha'}(\beta | \tilde{\beta}) = \frac{1}{{(\pi \alpha')}^{n-3}} \left( m(\beta | \tilde{\beta}) + \mathcal{O}( \alpha') \right) ,
\ens
where $n$ in the number of particles. The new theory can be understood as a bi-adjoint scalar with additional interaction terms that vanish in the infinite tension limit. As we will see, however, the precise knowledge of these terms is not necessary to compute the amplitudes. Instead, we will employ graphical rules that compute them exactly in $\alpha'$.

Before explaining the rules, let us provide some examples of $m_{\alpha'}(\beta | \tilde{\beta})$ that are useful to keep in mind. From the direct inversion of the string KLT kernel we obtain:
\be\label{eq:m3}
m_{\alpha'}(123|123) = 1,
\en
\be\label{eq:m4}
m_{\alpha'}(1234|1234) = \frac{1}{\tan (\pi \alpha' s_{12})} + \frac{1}{\tan (\pi \alpha' s_{23})}, \quad m_{\alpha'}(1234|1243) = -\frac{1}{\sin (\pi \alpha' s_{12})},
\en
\bes
m_{\alpha'}(12345|13254) = \frac{1}{\sin(\pi \alpha' s_{23}) \sin(\pi \alpha' s_{45})}, &&\; m_{\alpha'}(12435|14253) = \frac{1}{\sin(\pi \alpha' s_{24}) \sin(\pi \alpha' s_{35})},\tr\tr
m_{\alpha'}(12345|14253) = 0, &&\; m_{\alpha'}(12435|13254) = 0.\phantom{\frac{1}{1}}
\ens
As before, we use the notation $s_{a_1 a_2 \cdots a_m} = \sum_{1 \leq i<j \leq m} p_{a_i} \cdot p_{a_j}$. It is straightforward to see these amplitudes collapse to (\ref{eq:Example n=3}--\ref{eq:Example n=5}) in the $\alpha' \to 0$ limit. Since the Mandelstam invariants are always multiplied by the factor $\pi \alpha'$, from now on we will set $\pi \alpha' = 1$ for the sake of brevity. We also introduce the notation $\mathbb{I}_n$ to denote the identity permutation with $n$ labels. Then, some more interesting examples become:
\be
m_{\alpha'}(\mathbb{I}_6|126435) = \frac{1}{\sin s_{12} \,\sin s_{34} \,\sin s_{345}},
\en
\be
m_{\alpha'}(\mathbb{I}_6|126345) = \frac{1}{\sin s_{12} \,\sin s_{345}} \left(\frac{1}{\tan s_{34}} + \frac{1}{\tan s_{45}} \right),
\en
\be
m_{\alpha'}(\mathbb{I}_7|1276345) = \frac{1}{\sin s_{12}\,\sin s_{345}} \left( \frac{1}{\tan s_{34}} + \frac{1}{\tan s_{45}}\right) \left( \frac{1}{\tan s_{67}} + \frac{1}{\tan s_{712}}\right).
\en
These amplitudes factorize into products of inverse sine factors times the amplitude $m_{\alpha'}(\mathbb{I}_4 | \mathbb{I}_4 )$ from \eqref{eq:m4}. As we will find out, in general any off-diagonal amplitude, i.e., one with $\beta \neq \tilde{\beta}$, can be reduced into a product of the diagonal ones, $m_{\alpha'}(\beta|\beta)$, connected by $1/\sin (\pi \alpha' s_e)$ propagators. This motivates a separate discussion of the two cases.

For convenience, we have attached a \texttt{Mathematica} notebook as an ancillary file to this \texttt{arXiv} submission, which allows to reconstruct all the objects $m_{\alpha'}(\beta| \tilde{\beta})$ studied in this work, using the diagrammatic rules described below.

\subsection{\label{sec:Off-diagonal amplitudes}Off-diagonal Amplitudes}

We can compute the off-diagonal terms, i.e., $\beta \neq \tilde{\beta}$, by adapting the diagrammatic rules of the usual bi-adjoint theory \cite{Cachazo:2013iea}. It provides a convenient way of listing all the graphs allowed in both permutations at the same time. The procedure is as follows. First, draw points on a circle according to the permutation $\beta$. Next, connect the points with a loop joining them with respect to the other ordering, $\tilde{\beta}$. As a result, we obtain a set of polygons defined by the interior of this loop. For instance, for the examples given above, $m_{\alpha'}(\mathbb{I}_6|126435)$, $m_{\alpha'}(\mathbb{I}_6|126345)$ and $m_{\alpha'}(\mathbb{I}_7|1276345)$, we have respectively:
\be
\parbox[c]{6.5em}{\includegraphics[scale=.5]{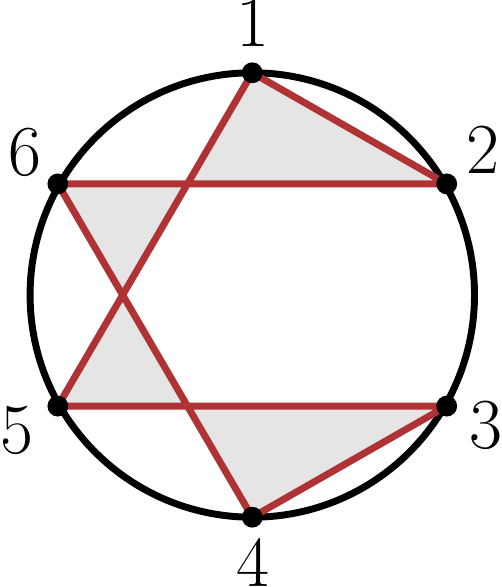}} \; ,\qquad \parbox[c]{6.5em}{\includegraphics[scale=.5]{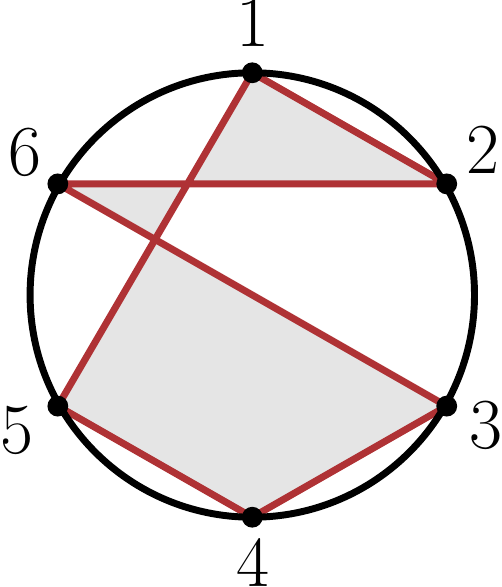}}\quad\, \text{and} \quad \parbox[c]{6.5em}{\includegraphics[scale=.5]{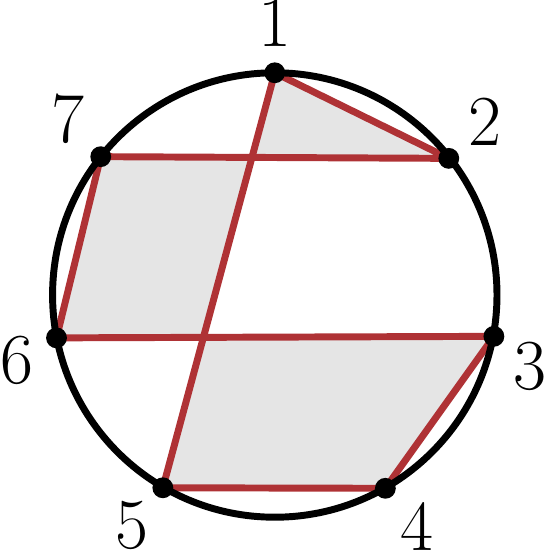}}\quad.
\en
Here, the permutations $\beta$ and $\tilde{\beta}$ are drawn with black and red lines respectively. The polygons are shown in grey. Starting from these diagrams, the prescription for obtaining the amplitude is very simple. Each shaded polygon corresponds to a sub-amplitude, and each vertex of the polygons corresponds to a single leg. As a result, this recipe has given us all the diagrams that are planar with respect to both $\beta$ and $\tilde{\beta}$ \cite{Cachazo:2013iea}.

Let us now specialize to the theory under consideration. We find the following rules. Internal legs are decorated with propagators of the form $1/\sin (\pi \alpha' s_{e})$, where $s_e = p_e^2/2$ is the norm of the momentum flowing through the leg $e$. In this way, we have introduced an infinite number of simple poles at $\alpha' s_e = 0, \pm1, \pm2, \ldots$ for each of the internal states. This means that if we choose to interpret the new object as a physical theory, it would contain massless, massive and tachyonic states. Note that in the $\alpha' \to 0$ limit we obtain the usual massless propagator $1/s_e$, which is consistent with the field theory bi-adjoint scalar.

Let us now dissect each diagram in turn. Firstly,
\be\label{eq:m-123456-126435}
m_{\alpha'}(\mathbb{I}_6|126435)\, = \parbox[c]{6.5em}{\includegraphics[scale=.5]{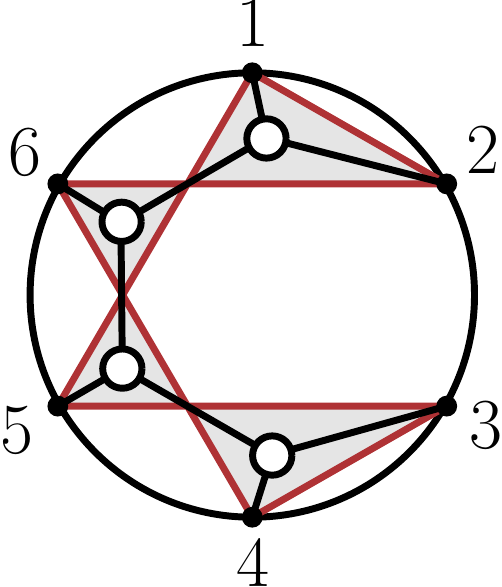}} =\, \frac{1}{\sin s_{12} \,\sin s_{34} \,\sin s_{345}}.
\en
In this example, all the sub-amplitudes, here denoted with white circles, are trivalent. From \eqref{eq:m3} we now that each of them evaluates to $1$. Therefore, the amplitude under consideration is just a product of propagators given by sines. Secondly,
\bes\label{eq:m-123456-126345}
m_{\alpha'}(\mathbb{I}_6|126345)\, = \parbox[c]{6.5em}{\includegraphics[scale=.5]{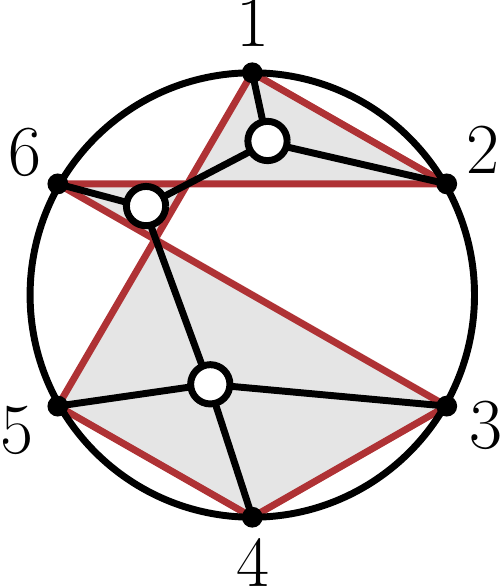}} &=&\, -\frac{1}{\sin s_{12} \,\sin s_{612}} \times\!\!\! \parbox[c]{6.5em}{\includegraphics[scale=.5]{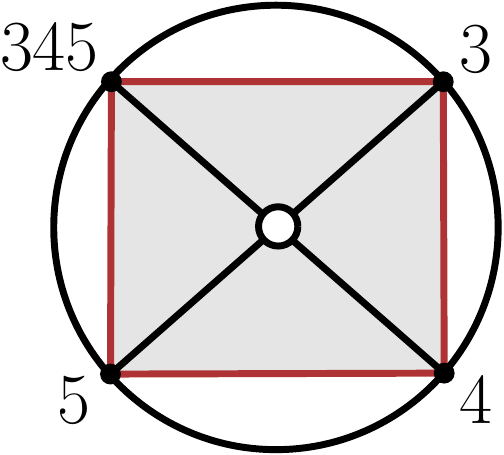}}\tr
&=&\, -\frac{1}{\sin s_{12} \,\sin s_{612}} \left(\frac{1}{\tan s_{34}} + \frac{1}{\tan s_{45}} \right).\qquad\quad
\ens
The amplitude can be written as a product of the propagators times the sub-amplitudes, two of which are trivial. In the last equality we have used the diagonal four-point amplitude given in \eqref{eq:m4}. We will explain the origin of the minus sign shortly. Finally, we have:
\bes\label{eq:m-1234567-1276345}
m_{\alpha'}(\mathbb{I}_7|1276345) &=& \parbox[c]{6.5em}{\includegraphics[scale=.5]{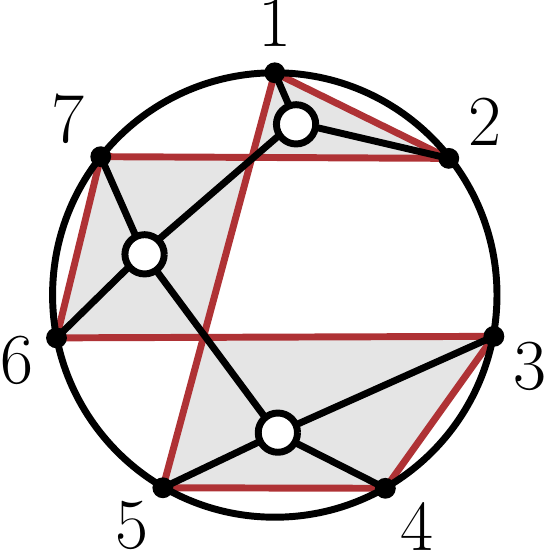}}\tr
&=& \frac{1}{\sin s_{12}\,\sin s_{345}} \,\times\!\! \parbox[c]{6.5em}{\includegraphics[scale=.5]{figures/m-123456-126345-small}}\;\;\times\;\, \parbox[c]{6.5em}{\includegraphics[scale=.5]{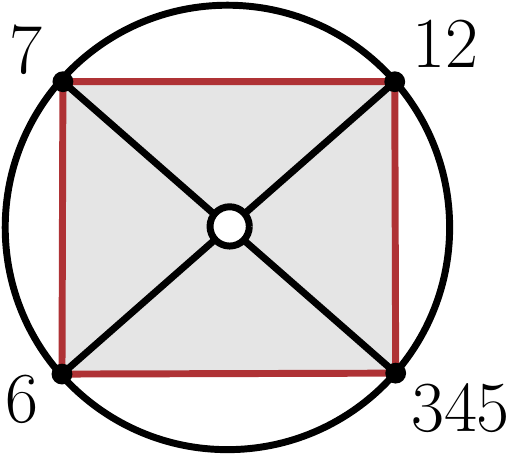}} \tr
&=& \frac{1}{\sin s_{12}\,\sin s_{345}} \left( \frac{1}{\tan s_{34}} + \frac{1}{\tan s_{45}}\right) \left( \frac{1}{\tan s_{67}} + \frac{1}{\tan s_{712}}\right).\qquad
\ens
Here we have used \eqref{eq:m4} twice. 

The overall signs of the amplitudes will be important later on. They are given in terms of the relative winding number between the two permutations, $w(\beta | \tilde{\beta})$. The rule is to first draw the permutation $\beta$ on a circle, and then follow the points according to the other permutation $\tilde{\beta}$ by always going clockwise. The relative winding number, $w(\beta | \tilde{\beta})$ is then given by the total number of cycles completed. For instance, for the above example \eqref{eq:m-123456-126345} we have:
\be\label{eq:winding}
w(\mathbb{I}_6 | 126345)\, = \parbox[c]{6.5em}{\includegraphics[scale=.5]{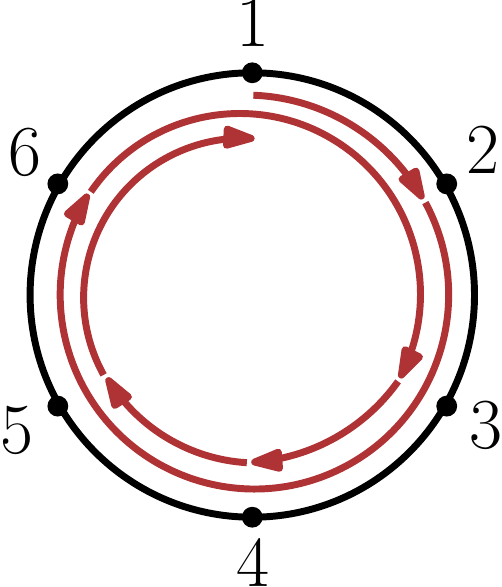}}\, =\, 2.
\en
The sign of the amplitude $m_{\alpha'}(\beta | \tilde{\beta})$ is then simply given by $(-1)^{w(\beta | \tilde{\beta}) + 1}$. We give a proof of this statement in appendix \ref{sec:Proof of the Sign}. In the above example, this prescription leads to a minus sign. Similarly, both \eqref{eq:m-123456-126435} and \eqref{eq:m-1234567-1276345} have winding number $3$, thus yielding a plus sign.

The above method allows for a convenient reduction of any off-diagonal amplitude to a product of propagators and the diagonal sub-amplitudes. Diagonal amplitudes, $m_{\alpha'}(\beta | \beta)$, are the ones containing the information about the $\alpha'$ corrections. Before we proceed to explaining how to compute them, let us comment on the vanishing amplitudes. When it is impossible to construct a tree-level graph compatible with two partial orderings, the amplitude necessarily vanishes. In the diagrammatic language, it means that the duals of the polygons form a loop.\footnote{Note, however, that the decomposition into polygons depends on the placement of labels on the outer circle. If a tree-level decomposition exists, it is unique \cite{Cachazo:2013iea}. If it does not, the corresponding amplitude vanishes. The trick is to always place the labels that are neighbouring in both orderings at infinitesimal separation \cite{Cachazo:2013iea}. For instance, in the last diagram of \eqref{eq:m-vanishing}, we have placed labels $1$ and $2$ very close to each other.} We have, for example:
\be\label{eq:m-vanishing}
\parbox[c]{6.5em}{\vspace{-.5em}\includegraphics[scale=.5]{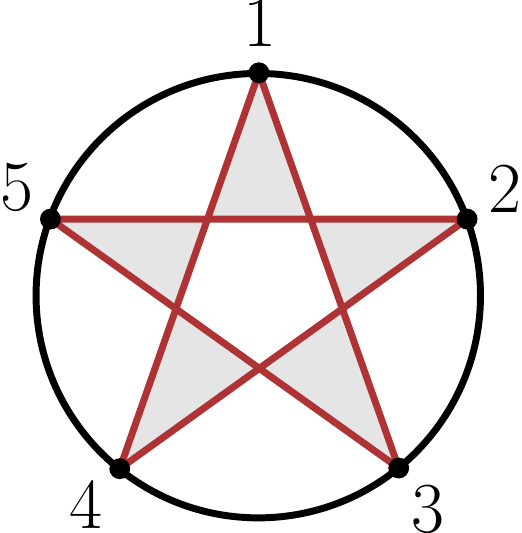}} \;\;=\; \parbox[c]{6.5em}{\includegraphics[scale=.5]{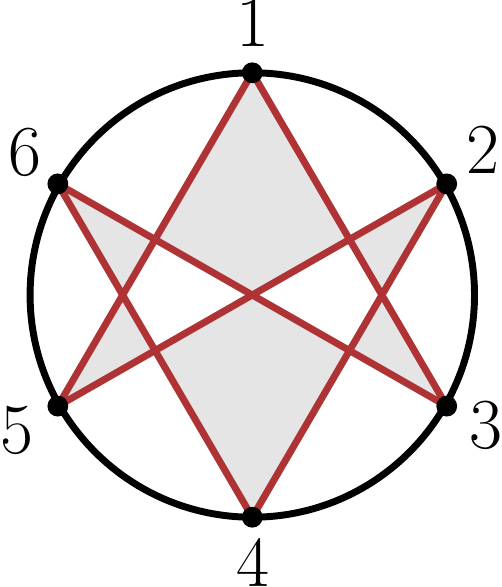}} \;\;=\; \parbox[c]{6.5em}{\includegraphics[scale=.5]{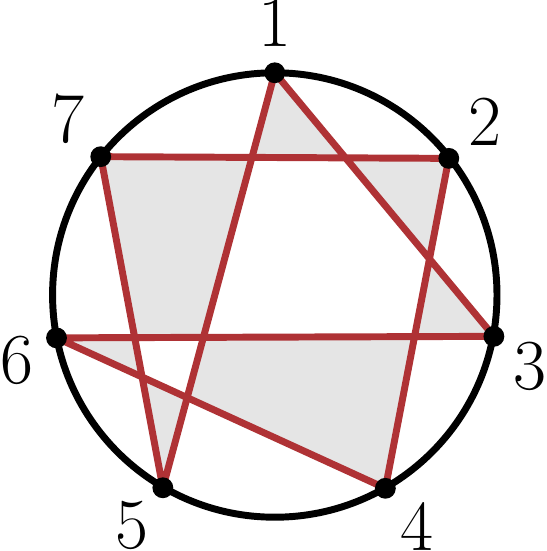}} \;\;\;=\; \parbox[c]{6.5em}{\includegraphics[scale=.5]{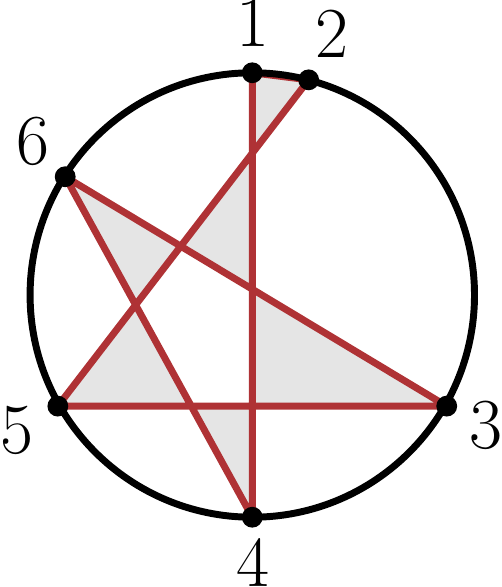}}\; =\; 0.
\en
In all the cases, it is impossible to find a tree-level graph consistent with planar embeddings $\beta$ and $\tilde{\beta}$. Vanishing amplitudes will be exploited later on, in order to construct matrices with zero entries, which are easier to invert.

\subsection{Diagonal Amplitudes}

The diagonal amplitudes, $m_{\alpha'}(\beta | \beta)$, contain the whole information about the higher-order interactions in this theory. Even without the full knowledge of their precise form, we will see that the on-shell amplitudes have a simple form amenable to a diagrammatic expansion. Without loss of generality we can focus on the identity permutations, i.e., $\beta= \tilde{\beta} = \mathbb{I}_n$. So far we have met the diagonal amplitudes \eqref{eq:m3} and \eqref{eq:m4}. Let us rewrite them as follows:
\be
m_{\alpha'}(\mathbb{I}_3 | \mathbb{I}_3 )\, = \parbox[c]{6.5em}{\vspace{-.5em}\includegraphics[scale=.5]{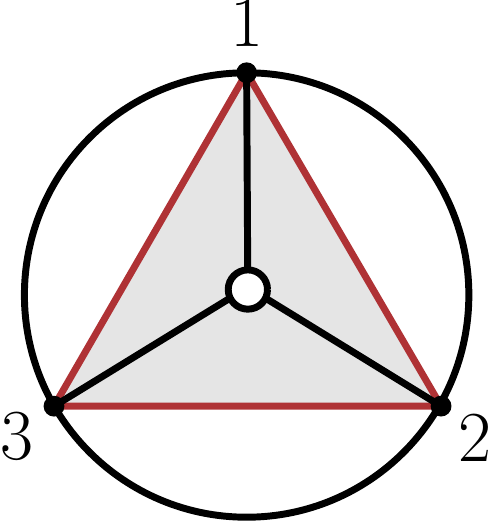}} = \parbox[c]{6em}{\vspace{-1em}\includegraphics[scale=.5]{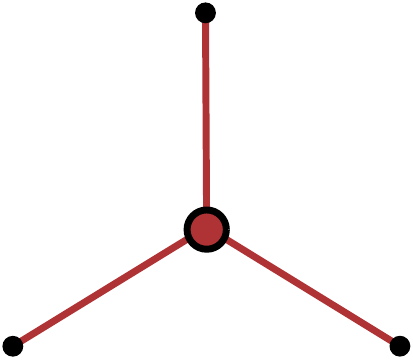}}\!\!\! =\; 1,
\en
\be\label{eq:m-1234-1234}
m_{\alpha'}(\mathbb{I}_4 | \mathbb{I}_4 )\, =\; \parbox[c]{6.5em}{\includegraphics[scale=.5]{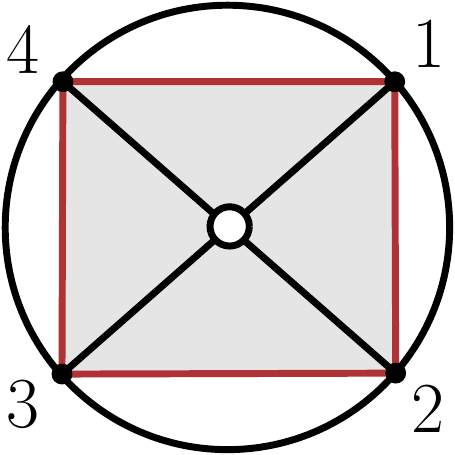}} =\; \parbox[c]{5em}{\includegraphics[scale=.5]{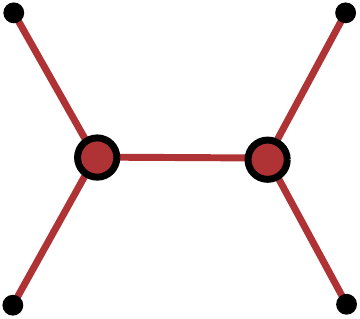}} \;+\;\, \parbox[c]{5em}{\includegraphics[scale=.5]{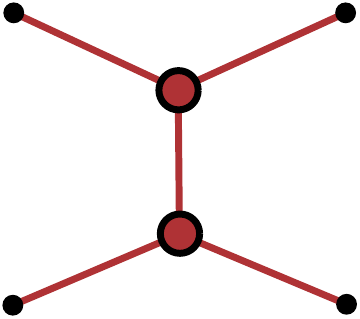}}\;
=\, \frac{1}{\tan s_{12}}\; +\; \frac{1}{\tan s_{23}}.
\en
Here, we have introduced a new expansion that is not related to the previous sub-section. Of course, the $m_{\alpha'}(\beta | \tilde{\beta})$ amplitudes could still be written in the expansion of sine propagators plus additional interaction terms. The reason for introducing a different set of rules is that the amplitudes turn out to have a very simple expressions in this new language. This time, each red internal line, $e$, corresponds to a propagator of the form $1/\tan (\pi \alpha' s_e)$. Hence, again, it produces an infinite number of simple poles at $\alpha' s_e = 0, \pm1, \pm2, \ldots$. Each trivalent red vertex carries the same factor as before, $1$. It turns out that these graphical rules extend to higher multiplicities.

Let us see how they work for the next case, $n=5$:
\bes
m_{\alpha'}(\mathbb{I}_5 | \mathbb{I}_5 )\, = \parbox[c]{6.5em}{\vspace{-.5em}\includegraphics[scale=.5]{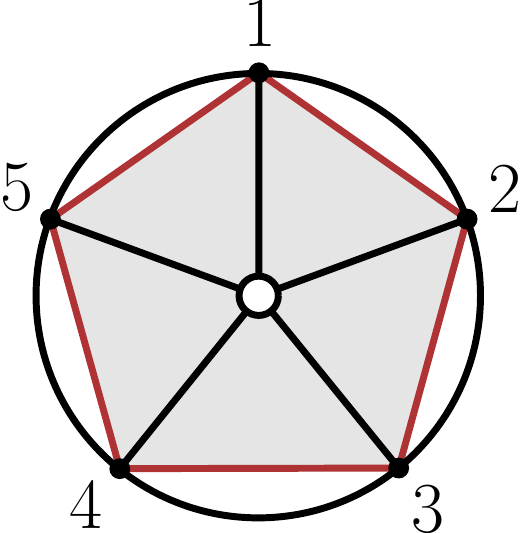}}\, &=&\, \parbox[c]{6em}{\includegraphics[scale=.5]{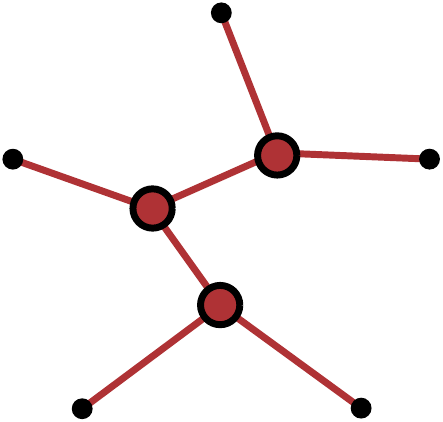}}\, + \,\parbox[c]{6em}{\includegraphics[scale=.5]{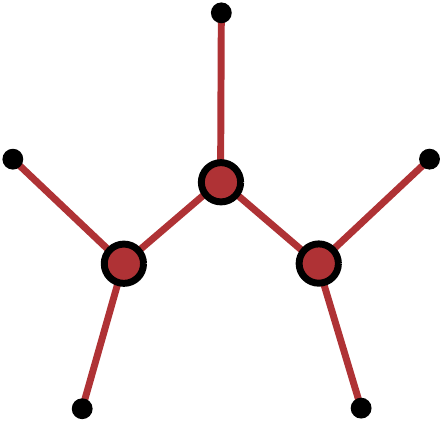}}\, +\, \parbox[c]{6em}{\includegraphics[scale=.5]{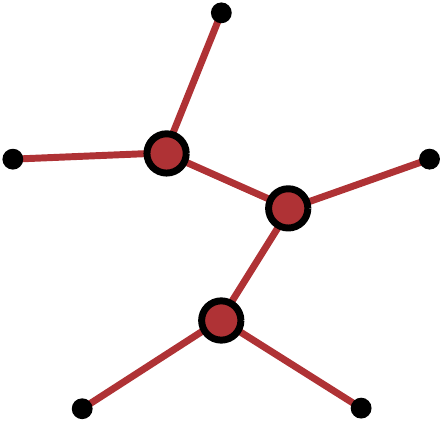}} \tr
&&+\, \parbox[c]{6em}{\includegraphics[scale=.5]{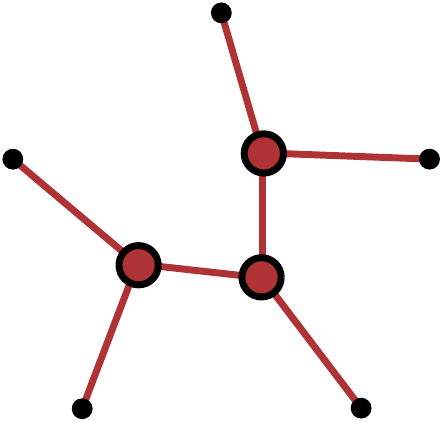}}\, +\, \parbox[c]{6em}{\includegraphics[scale=.5]{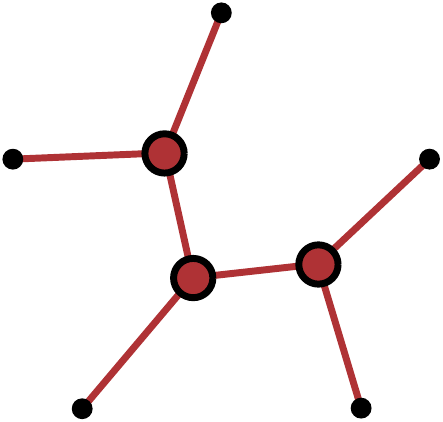}} \,+\, \parbox[c]{6em}{\includegraphics[scale=.5]{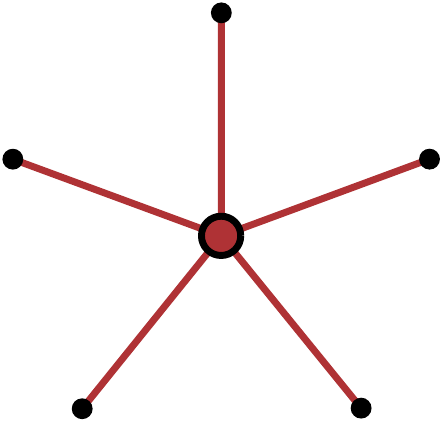}} \tr\tr
&=& \frac{1}{\tan s_{12}\, \tan s_{34}} + \frac{1}{\tan s_{23}\, \tan s_{45}} + \frac{1}{\tan s_{34}\, \tan s_{51}} \tr \vspace{-1em}\tr
&& + \frac{1}{\tan s_{45}\, \tan s_{12}} + \frac{1}{\tan s_{51}\, \tan s_{23}} + 1.
\ens
The first five terms are produced by trivalent graphs with the same vertex as before. These terms are fixed by requiring that the amplitude factorizes correctly on all of the massless poles. There is also a contribution that stays finite on all the factorization channels. It takes a very simple form of a contact term carrying a dimensionless factor $1$. Note that in the infinite tension limit $\alpha' \to 0$, the contact term does not contribute and we recover the usual bi-adjoint scalar amplitude. We continue with $n=6$:
\bes
m_{\alpha'}(\mathbb{I}_6 | \mathbb{I}_6 ) &=& \left(\; \parbox[c]{4em}{\includegraphics[scale=.4]{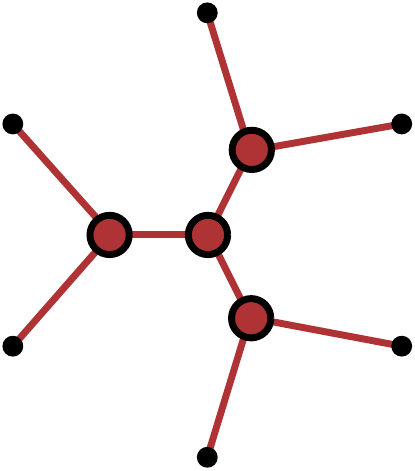}} \;\;+\; \parbox[c]{4em}{\text{13 other}\\ \text{\;\, terms}} \right) + \left(\; \parbox[c]{4em}{\includegraphics[scale=.4]{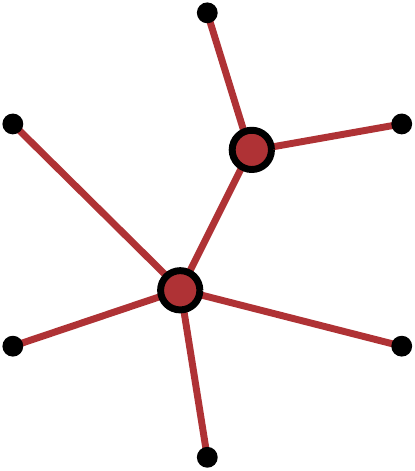}} \;\;+\; \parbox[c]{3.5em}{\text{5 other}\\ \text{ terms}}\right)\tr
&=& \left( \frac{1}{\tan s_{12}\, \tan s_{34}\, \tan s_{56}} + \;\cdots\; \right)+ \left( \frac{1}{\tan s_{12}} + \;\cdots\; \right).
\ens
As we can see, the presence of the new five-valent vertex introduced a hierarchy in the expansion. The terms in the first bracket go as $\alpha'^{-3}$ close to the field theory limit, while the second bracket behaves as $\alpha'^{-1}$. Because of this, only the first family survives the $\alpha' \to 0$ limit, as expected. There are no new vertices introduced at this stage.

Now that the general rule is established, we want to find out the form of the additional contact terms appearing at higher multiplicities. We find:
\be
m_{\alpha'}(\mathbb{I}_7 | \mathbb{I}_7 ) =
\left(\; \parbox[c]{4em}{\vspace{-.0em}\includegraphics[scale=.4]{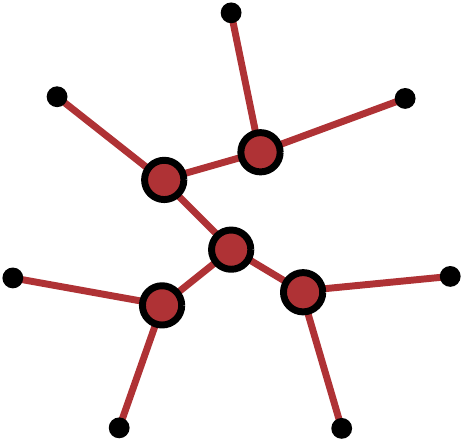}} \;\;\;\;+\; \parbox[c]{4em}{\text{41 other}\\ \text{\;\, terms}} \right) + \left( \parbox[c]{4em}{\vspace{-.0em}\includegraphics[scale=.4]{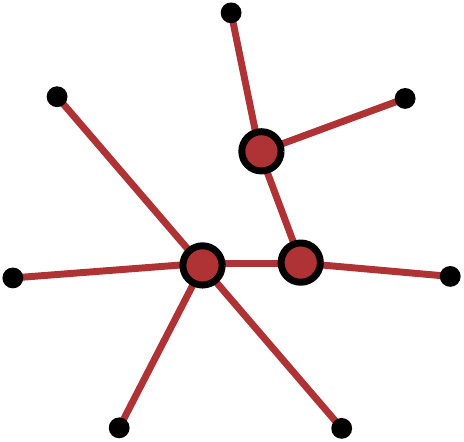}} \;\;\;\;+\; \parbox[c]{4em}{\text{27 other}\\ \text{\;\, terms}} \right)\nn
+\; \parbox[c]{4em}{\vspace{-.0em}\includegraphics[scale=.4]{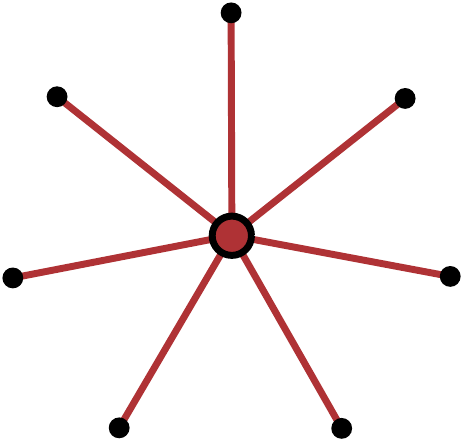}}\;\;\;\;\;,
\en
where the last term is a seven-valent vertex equal to $2$. Here, we have again sorted the terms depending on their leading $\alpha'$ order. Finally:
\bes
m_{\alpha'}(\mathbb{I}_8 | \mathbb{I}_8 ) &=& \left(\; \parbox[c]{4em}{\includegraphics[scale=.4]{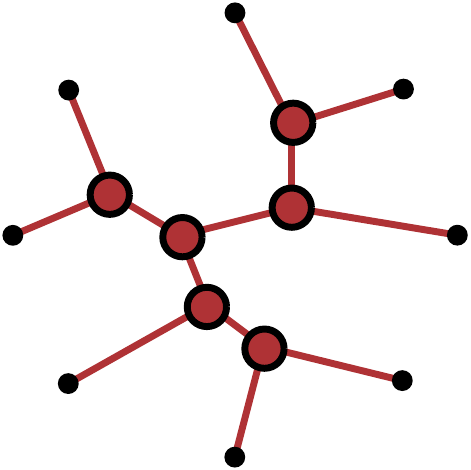}} \;\;\;\;\;+\; \parbox[c]{4.5em}{\text{131 other}\\ \text{\;\;\, terms}} \right) + \left(\; \parbox[c]{4em}{\includegraphics[scale=.4]{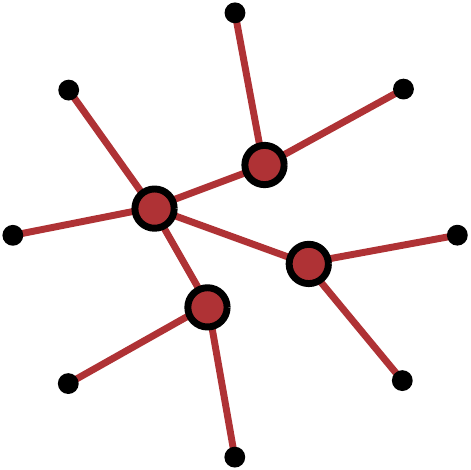}} \;\;\;\;\;+\; \parbox[c]{4.5em}{\text{123 other}\\ \text{\;\, terms}} \right) \tr
&& + \left(\; \parbox[c]{4em}{\includegraphics[scale=.4]{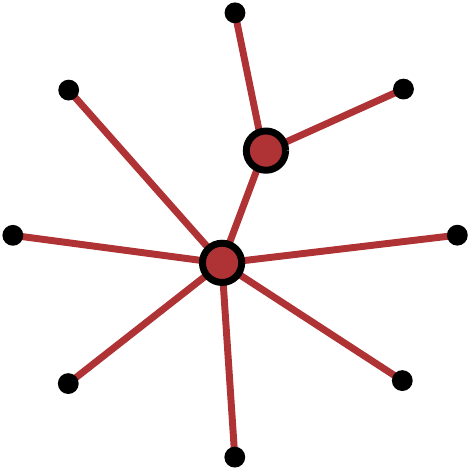}} \;\;\;\;\;+\; \parbox[c]{3.5em}{\text{7 other}\\ \text{\, terms}} \right),\qquad
\ens
\bes
m_{\alpha'}(\mathbb{I}_9 | \mathbb{I}_9 ) &=& \left(\; \parbox[c]{4em}{\includegraphics[scale=.4]{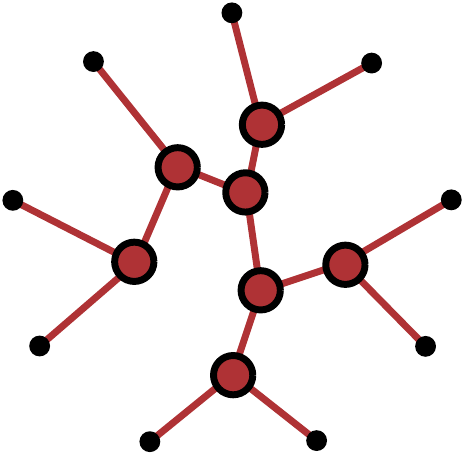}} \;\;\;\;\;+\; \parbox[c]{4.5em}{\text{428 other}\\ \text{\;\;\, terms}} \right) + \left(\; \parbox[c]{4em}{\includegraphics[scale=.4]{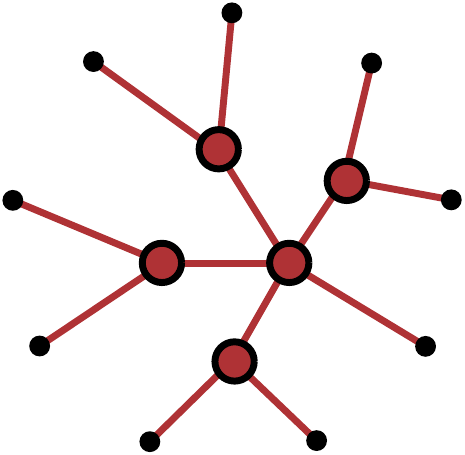}} \;\;\;\;\;+\; \parbox[c]{4.5em}{\text{530 other}\\ \text{\;\;\, terms}} \right)\tr
&& + \left(\; \parbox[c]{4em}{\includegraphics[scale=.4]{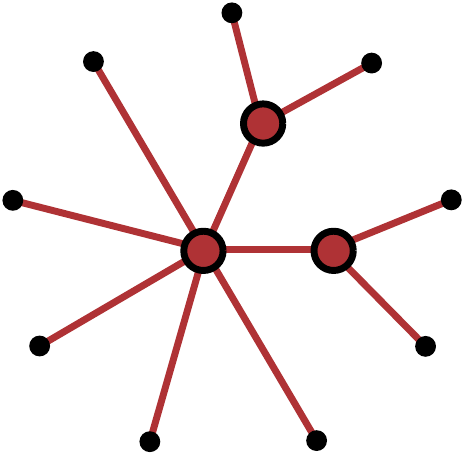}} \;\;\;\;\;+\; \parbox[c]{4em}{\text{53 other}\\ \text{\;\, terms}} \right) +\; \parbox[c]{4em}{\includegraphics[scale=.4]{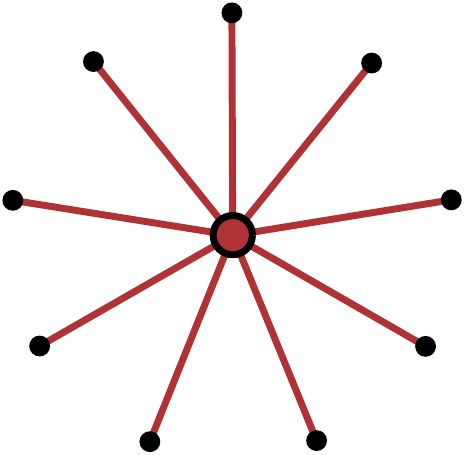}}\;\;\;\;\;.
\ens
We find that the last nine-valent vertex carries a factor $5$. To summarize, we have found that the diagonal amplitudes $m_{\alpha'}(\beta | \beta)$ can be calculated from a graph expansion, where each propagator carries an inverse tangent factor, and $3,5,7,9$-valent vertices come with factors $1,1,2,5$ respectively. This pattern seems to generate Catalan numbers. It is tempting to conjecture that all the remaining vertices follow this pattern, i.e.,
\be
\parbox[c]{8em}{\includegraphics[scale=.5]{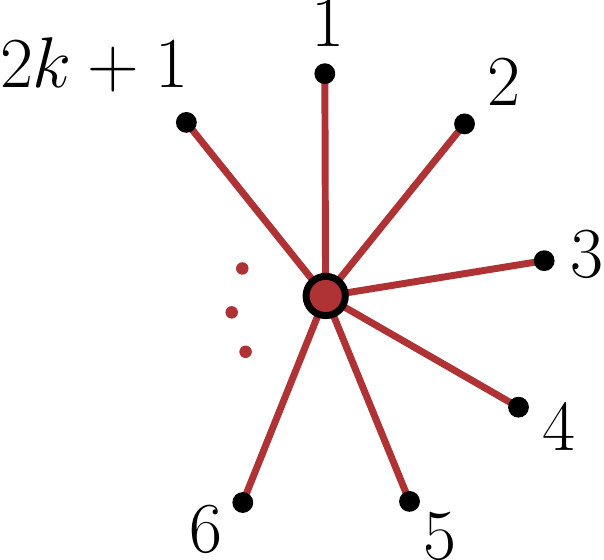}}\; = \; C_{k-1},
\en
where $C_k$ is the $k$-th Catalan number \cite{CatalanNumbers}.

\section{\label{sec:KLT Relations}KLT Relations}

KLT relations were originally derived in \cite{Kawai:1985xq} using the holomorphic factorization properties of the closed string amplitudes. The explicit formula for all multiplicities in the $\alpha' \to 0$ limit was given in \cite{Bern:1998sv} and later proven in a more general setting in \cite{BjerrumBohr:2010hn}. In this section, we show how to construct KLT relations from the objects $m_{\alpha'}(\beta | \tilde{\beta})$ introduced in this work. We have verified validity of this construction numerically for $n \leq 10$ and it remains a conjecture for higher multiplicities.

Using the graphical rules introduced in the previous section, we can calculate the KLT relations according to
\bes\label{eq:string KLT2}
\mathcal{M}^{\text{closed}}_n = \mathcal{A}^{\text{open}}(\beta) \; m_{\alpha'}^{-1}(\beta | \tilde{\beta})\; \mathcal{A}^{\text{open}}(\tilde{\beta}).
\ens
Here, we treat $m_{\alpha'}(\beta | \tilde{\beta} )$ as an $(n-3)! \times (n-3)!$ matrix with columns and rows labelled by the permutations $\beta$ and $\tilde{\beta}$. There is no restriction on the permutations we use for the open string partial amplitudes, as long as they form an independent basis, so that $m_{\alpha'}(\beta | \tilde{\beta} )$ is an invertible matrix.

Let us start with an illustrative example for $n=4$ and orderings $\beta = \tilde{\beta} = (1234)$. We can use the identity $\sin(\pi z) = \pi / \Gamma(z) \Gamma(1-z)$ to rewrite the amplitude \eqref{eq:m-1234-1234},
\bes
m_{\alpha'} (\mathbb{I}_4 | \mathbb{I}_4) &=& \frac{1}{\tan (\pi \alpha' s)} + \frac{1}{\tan (\pi \alpha' t)} = \frac{\sin(\pi \alpha' (s+t))}{\sin(\pi \alpha' s)\, \sin(\pi \alpha' t)}\tr
&=& \frac{\Gamma(\alpha' s)\, \Gamma(1- \alpha' s)\, \Gamma(\alpha' t)\, \Gamma(1- \alpha' t)}{\pi\, \Gamma(-\alpha' u)\, \Gamma(1+\alpha' u)}.
\ens
Here we are using the convention $s=s_{12}$, $t=s_{23}$, $u=s_{13}$, $s+t+u=0$. Using \eqref{eq:string KLT2} with the Veneziano amplitude \cite{Veneziano:1968yb} and working up to a universal kinematic factor we obtain:
\bes
\mathcal{M}_4^{\text{closed}} &=& \mathcal{A}^{\text{open}}(1234)\; m^{-1}_{\alpha'}(1234|1234)\; \mathcal{A}^{\text{open}}(1234) \tr
&=& \left( \frac{\Gamma(\alpha' s)\, \Gamma(\alpha' t)}{\Gamma(1+ \alpha' s+ \alpha' t)} \right) \left( \frac{\Gamma(\alpha' s)\, \Gamma(1- \alpha' s)\, \Gamma(\alpha' t)\, \Gamma(1- \alpha' t)}{\Gamma(-\alpha' u)\, \Gamma(1+ \alpha' u)} \right)^{-1} \left( \frac{\Gamma(\alpha' s)\, \Gamma(\alpha' t)}{\Gamma(1+\alpha' s+\alpha' t)} \right) \tr
&=& - \frac{\Gamma(\alpha' s)\, \Gamma(\alpha' t)\, \Gamma(\alpha' u)}{\Gamma(1-\alpha' s)\, \Gamma(1-\alpha' t)\, \Gamma(1-\alpha' u)},
\ens
which is the Virasoro-Shapiro amplitude \cite{Virasoro:1969me,Shapiro:1970gy}.

We can also use any other basis for the KLT expansion, for example:
\bes
\mathcal{M}_4^{\text{closed}} &=& \mathcal{A}^{\text{open}}(1234)\left(\; \parbox[c]{5.1em}{\includegraphics[scale=.4]{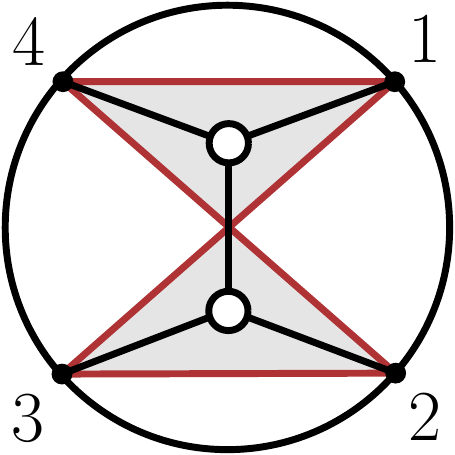}} \right)^{-1}\!\!\!\! \mathcal{A}^{\text{open}}(1324) \tr \vspace{-1em}\tr
&=& - \sin(\pi \alpha' t)\, \mathcal{A}^{\text{open}}(1234)\, \mathcal{A}^{\text{open}}(1324),
\ens
or alternatively,
\bes
\mathcal{M}_4^{\text{closed}} &=& \mathcal{A}^{\text{open}}(1234) \left(\; \parbox[c]{5.1em}{\includegraphics[scale=.4]{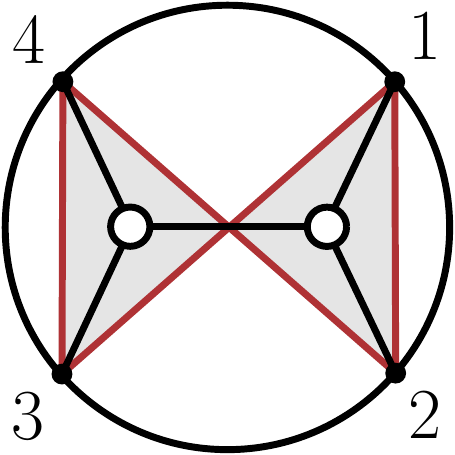}} \right)^{-1}\!\!\!\! \mathcal{A}^{\text{open}}(1243) \tr \vspace{-1em} \tr
&=& - \sin(\pi \alpha' s)\, \mathcal{A}^{\text{open}}(1234)\, \mathcal{A}^{\text{open}}(1243).
\ens

For $n=5$, the original KLT relation has permutations $\beta \in \left\{ (12345), (12435)\right\}$ and $\tilde{\beta} \in \left\{ (13254), (14253)\right\}$. Reinterpreted in our new language, the expression reads
\bes
\mathcal{M}_5^{\text{closed}} &=& \begin{bmatrix} \mathcal{A}^{\text{open}}(12345)\\ \mathcal{A}^{\text{open}}(12435) \end{bmatrix}^\intercal \begin{bmatrix} \;\,\parbox[c]{5.8em}{\includegraphics[scale=.4]{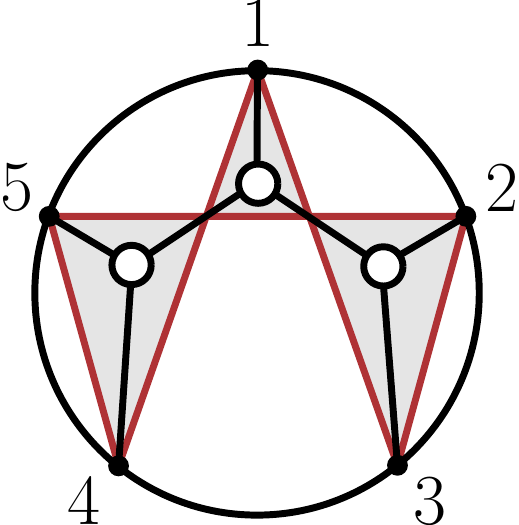}} & \parbox[c]{5.8em}{\includegraphics[scale=.4]{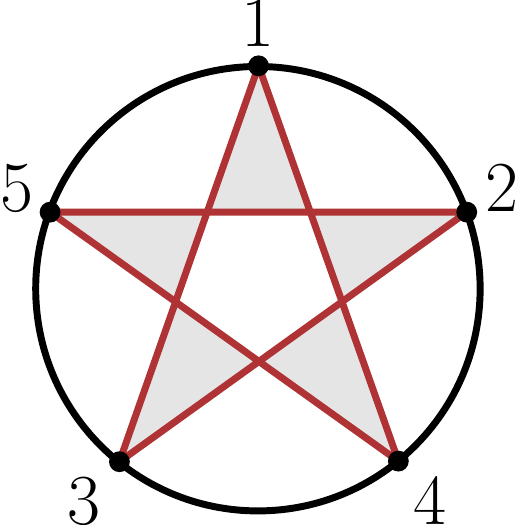}} \\ \vspace{-.9em}\\ \;\,\parbox[c]{5.8em}{\includegraphics[scale=.4]{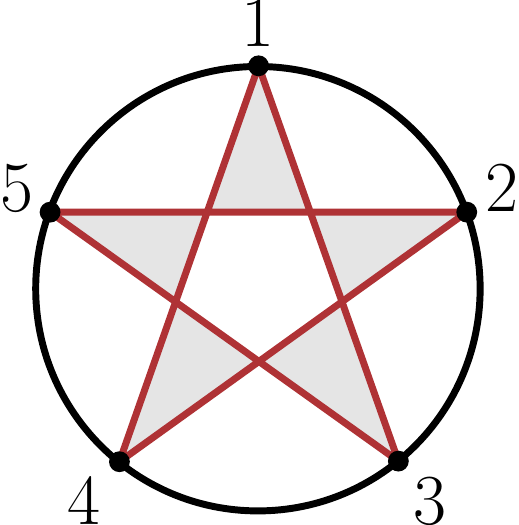}} & \parbox[c]{5.8em}{\includegraphics[scale=.4]{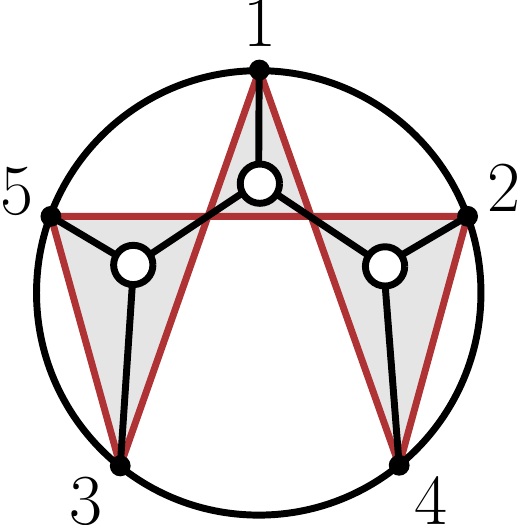}} \end{bmatrix}^{-1} \begin{bmatrix} \mathcal{A}^{\text{open}}(13254)\\ \mathcal{A}^{\text{open}}(14253) \end{bmatrix}\nn
\ens
\bes
&=& \begin{bmatrix} \mathcal{A}^{\text{open}}(12345)\\ \mathcal{A}^{\text{open}}(12435) \end{bmatrix}^\intercal \begin{bmatrix} \dfrac{1}{\sin s_{23} \, \sin s_{45}} & 0 \\ 0 & \dfrac{1}{\sin s_{24}\, \sin s_{35}} \end{bmatrix}^{-1} \begin{bmatrix} \mathcal{A}^{\text{open}}(13254)\\ \mathcal{A}^{\text{open}}(14253) \end{bmatrix}\tr\tr
&=& \sin (\pi \alpha' s_{23}) \sin (\pi \alpha' s_{45}) \,\mathcal{A}^{\text{open}}(12345)\, \mathcal{A}^{\text{open}}(13254) + (3 \leftrightarrow 4),
\ens
which correctly reduces to the field theory relation given in \eqref{eq:KLT5}. Note how the special choice of permutations gave a diagonal matrix, which is easier to invert. In general, the original KLT matrix will be block diagonal with each block of size $\lfloor \frac{n-3}{2} \rfloor! \lceil \frac{n-3}{2} \rceil! \times \lfloor \frac{n-3}{2} \rfloor! \lceil \frac{n-3}{2} \rceil!$. It is straightforward to see how the specific ordering chosen in \cite{Kawai:1985xq,Bern:1998sv} leads to the ``star'' configurations annihilating all the $\alpha'$-corrected bi-adjoint amplitudes outside of the diagonal blocks, as in the example above.

As an exercise, let us compute KLT relations for another choice of permutations, say $\beta \in \left\{ (13254), (14253)\right\}$ and $\tilde{\beta} \in \left\{ (12354), (12435)\right\}$. We then have:
\bes
\mathcal{M}_5^{\text{closed}} &=& \begin{bmatrix} \mathcal{A}^{\text{open}}(13254)\\ \mathcal{A}^{\text{open}}(14253) \end{bmatrix}^\intercal \begin{bmatrix} \;\,\parbox[c]{5.8em}{\includegraphics[scale=.4]{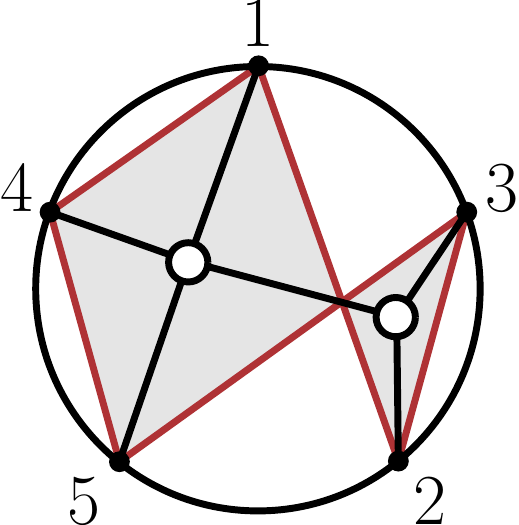}} & \parbox[c]{5.8em}{\includegraphics[scale=.4]{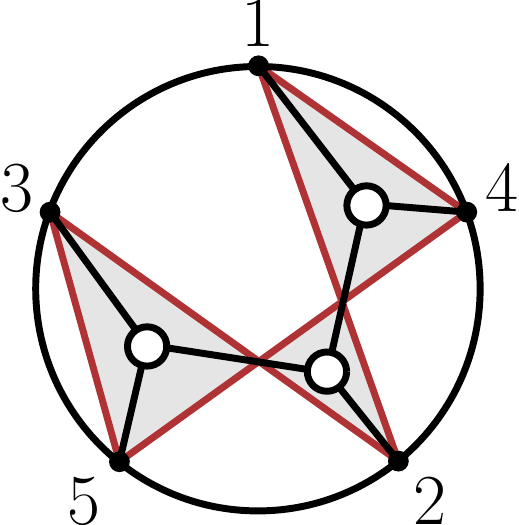}} \\ \vspace{-.9em}\\ \;\,\parbox[c]{5.8em}{\includegraphics[scale=.4]{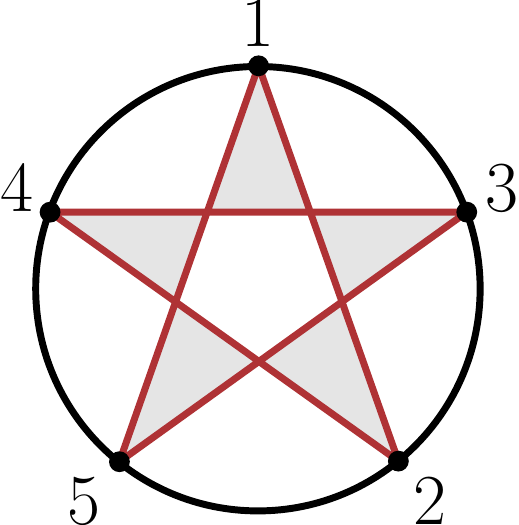}} & \parbox[c]{5.8em}{\includegraphics[scale=.4]{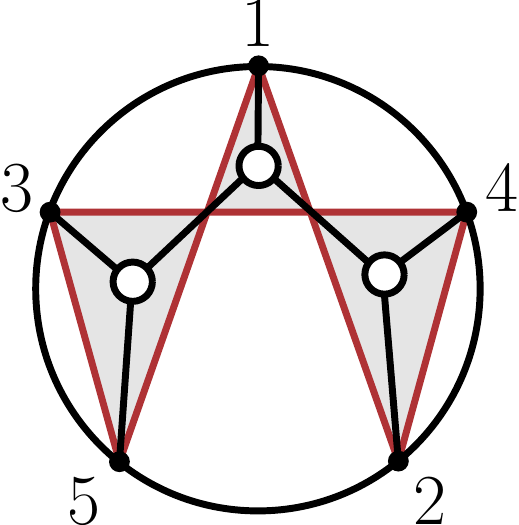}} \end{bmatrix}^{-1} \begin{bmatrix} \mathcal{A}^{\text{open}}(12354)\\ \mathcal{A}^{\text{open}}(12435) \end{bmatrix}\tr
&=& \begin{bmatrix} \mathcal{A}^{\text{open}}(13254)\\ \mathcal{A}^{\text{open}}(14253) \end{bmatrix}^\intercal \begin{bmatrix} -\dfrac{1}{\sin s_{23}}\left( \dfrac{1}{\tan s_{14}} + \dfrac{1}{\tan s_{45}}\right) & \dfrac{1}{\sin s_{14}\, \sin s_{35}} \\ 0 & \dfrac{1}{\sin s_{24}\, \sin s_{35}} \end{bmatrix}^{-1} \begin{bmatrix} \mathcal{A}^{\text{open}}(12354)\\ \mathcal{A}^{\text{open}}(12435) \end{bmatrix}\nn
\ens
\bes
=&& \sin (\pi \alpha' s_{24}) \sin (\pi \alpha' s_{35}) \mathcal{A}^{\text{open}}(12435) \mathcal{A}^{\text{open}}(14253)\tr
&& \;+ \frac{\sin (\pi \alpha' s_{23})\, \sin (\pi \alpha' s_{45})}{\sin (\pi \alpha' (s_{14} + s_{45}))}\, \mathcal{A}^{\text{open}}(13254) \Big( \sin (\pi \alpha' s_{24})\, \mathcal{A}^{\text{open}}(12435) \tr
&& \qquad\qquad\qquad\qquad\qquad\qquad\qquad\qquad\qquad - \sin (\pi \alpha' s_{14})\, \mathcal{A}^{\text{open}}(12354) \Big),
\ens
which can be verified using explicit formulae for the string amplitudes, e.g., \cite{Medina:2002nk}.

As the final example, we will reproduce the $n=6$ KLT kernel of \cite{Kawai:1985xq,Bern:1998sv}. In this case, the columns are labelled by $\beta \in \{ (123456),(124356),(132456),(134256),(142356),(143256)\}$ and the rows by $\tilde{\beta} \in \{(153462), (154362), (152463), (154263), (152364), (153264)\}$.

\noindent
After a tedious but straightforward calculation we obtain:
\be
m_{\alpha'}(\beta|\tilde{\beta}) = \begin{bmatrix} \;\parbox[c]{4em}{\includegraphics[scale=.3]{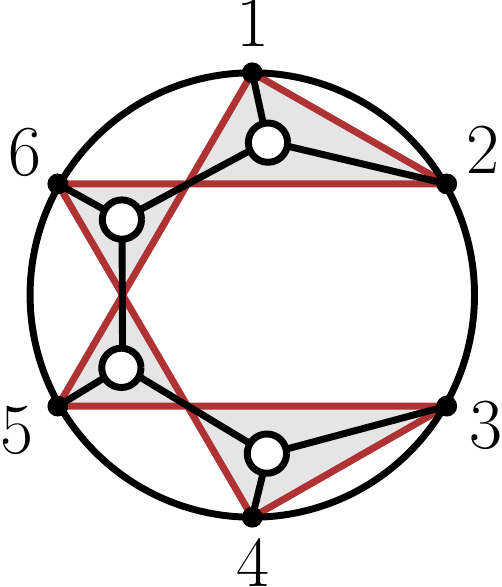}} & \parbox[c]{4em}{\includegraphics[scale=.3]{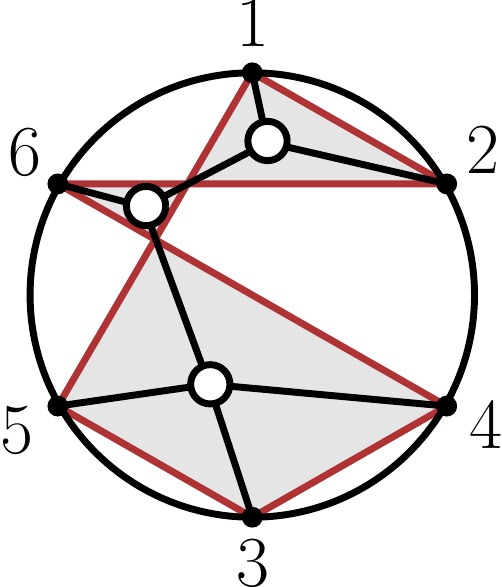}} & \parbox[c]{4em}{\includegraphics[scale=.3]{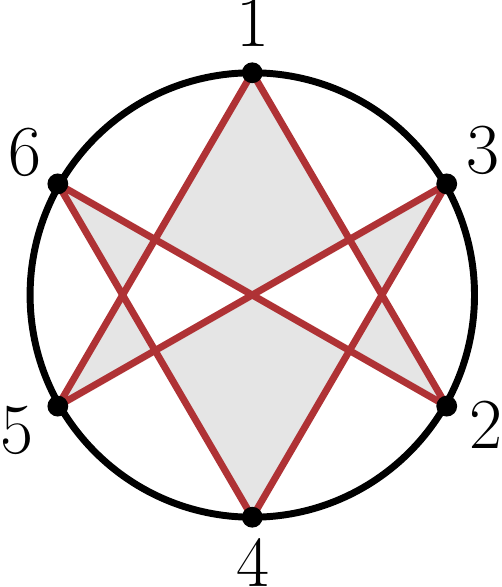}} & \parbox[c]{4em}{\includegraphics[scale=.3]{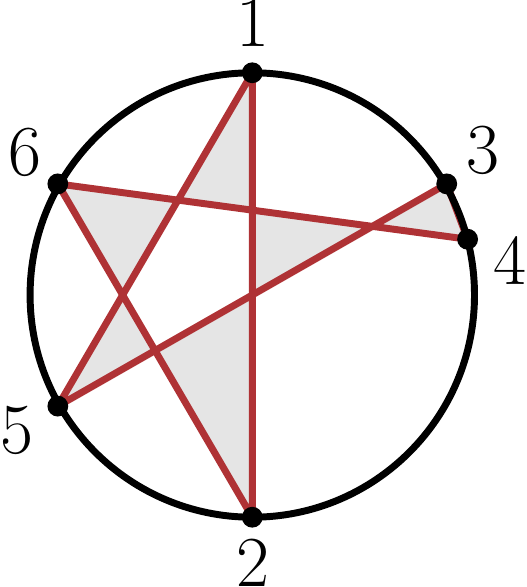}} & \parbox[c]{4em}{\includegraphics[scale=.3]{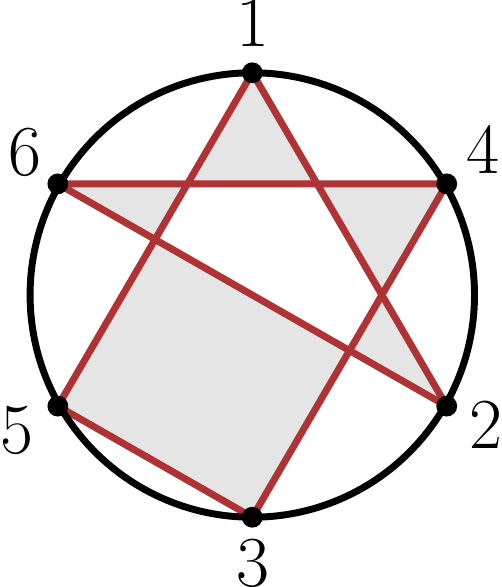}} &
\parbox[c]{4em}{\includegraphics[scale=.3]{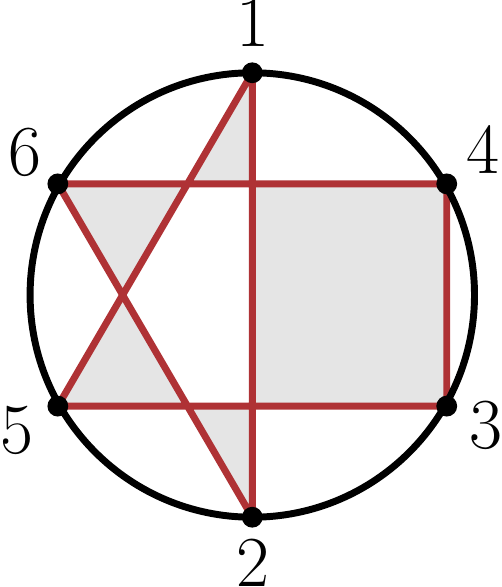}}\;\; \\ \vspace{-1em} \\
\parbox[c]{4em}{\includegraphics[scale=.3]{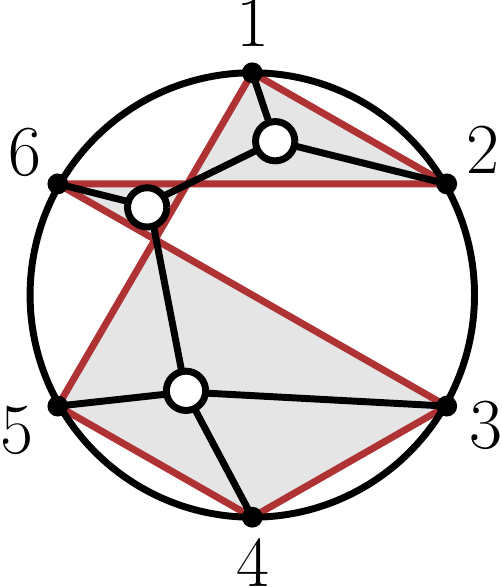}} & \parbox[c]{4em}{\includegraphics[scale=.3]{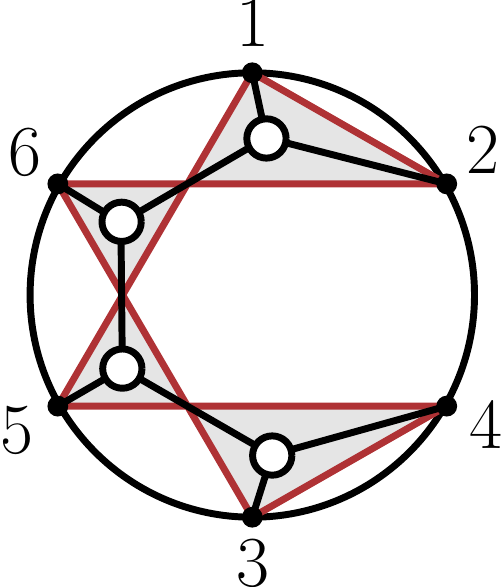}} & \parbox[c]{4em}{\includegraphics[scale=.3]{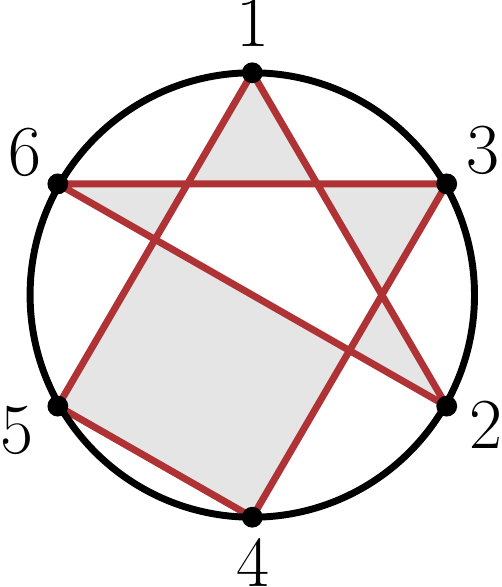}} & \parbox[c]{4em}{\includegraphics[scale=.3]{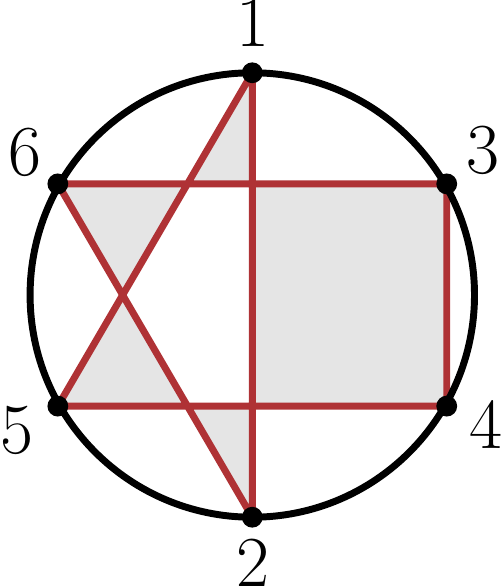}} & \parbox[c]{4em}{\includegraphics[scale=.3]{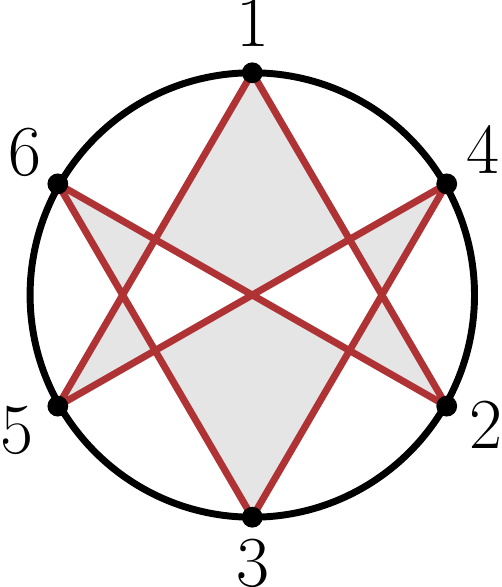}} &
\parbox[c]{4em}{\includegraphics[scale=.3]{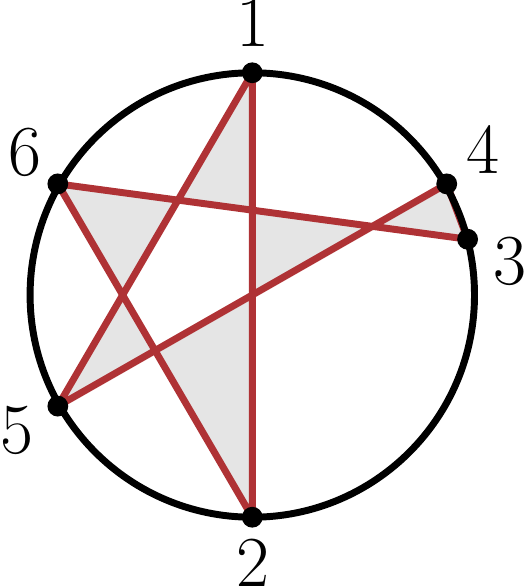}} \\ \vspace{-1em} \\
\parbox[c]{4em}{\includegraphics[scale=.3]{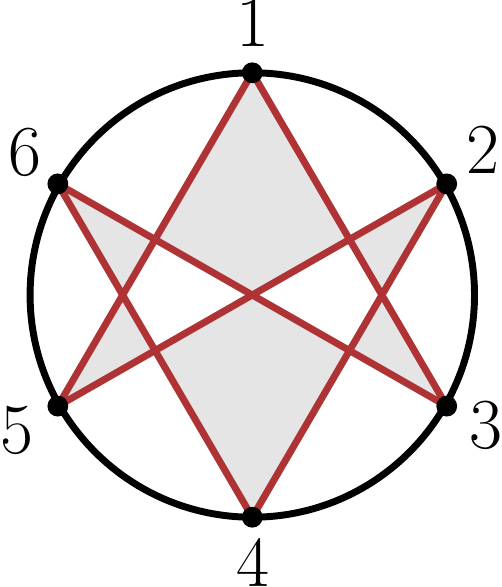}} & \parbox[c]{4em}{\includegraphics[scale=.3]{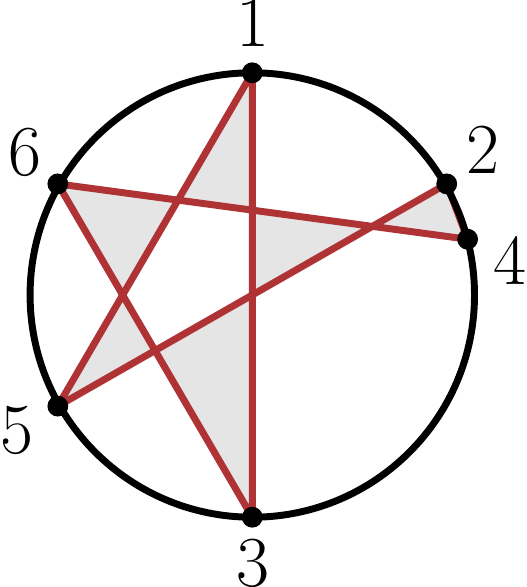}} & \parbox[c]{4em}{\includegraphics[scale=.3]{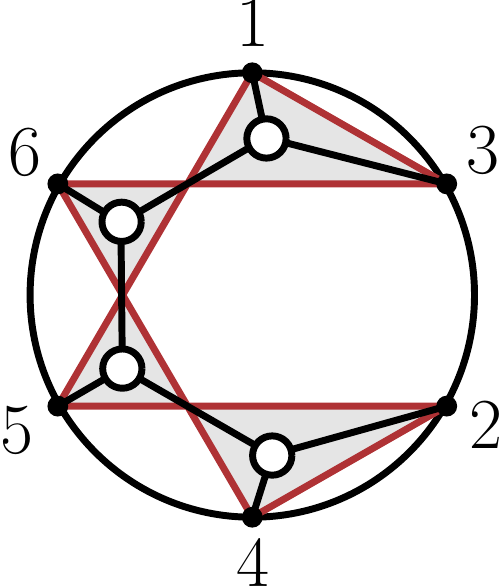}} & \parbox[c]{4em}{\includegraphics[scale=.3]{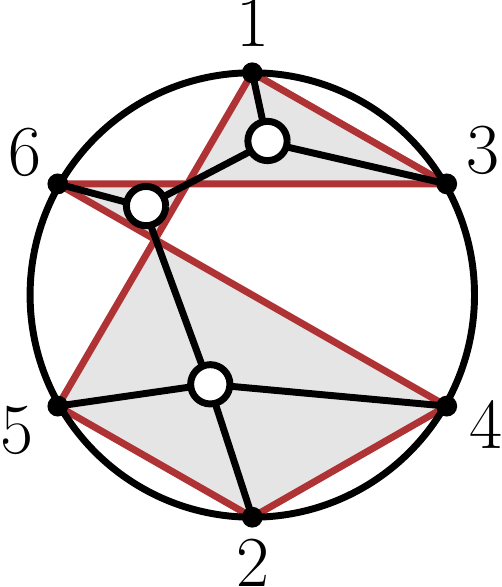}} & \parbox[c]{4em}{\includegraphics[scale=.3]{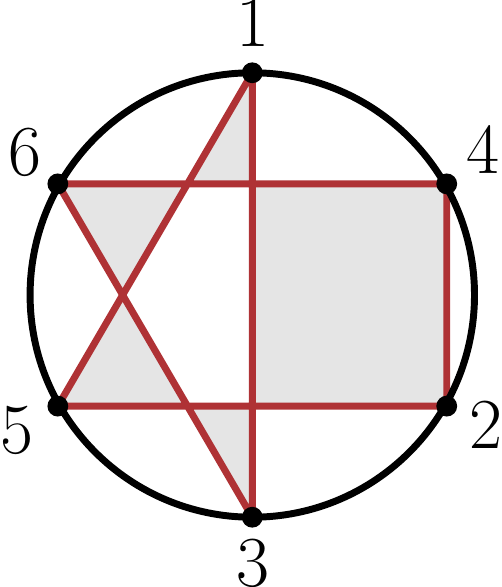}} &
\parbox[c]{4em}{\includegraphics[scale=.3]{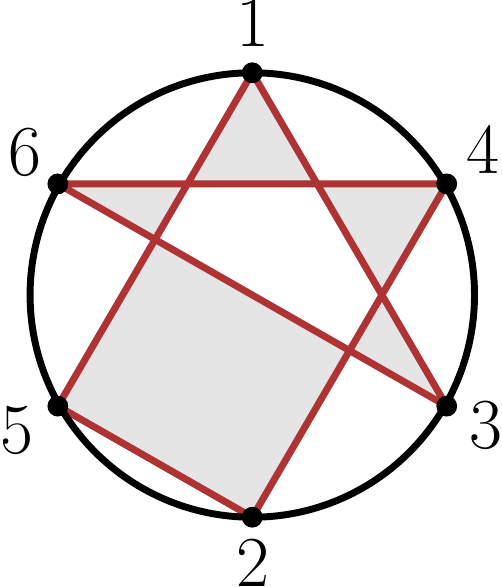}} \\ \vspace{-1em} \\
\parbox[c]{4em}{\includegraphics[scale=.3]{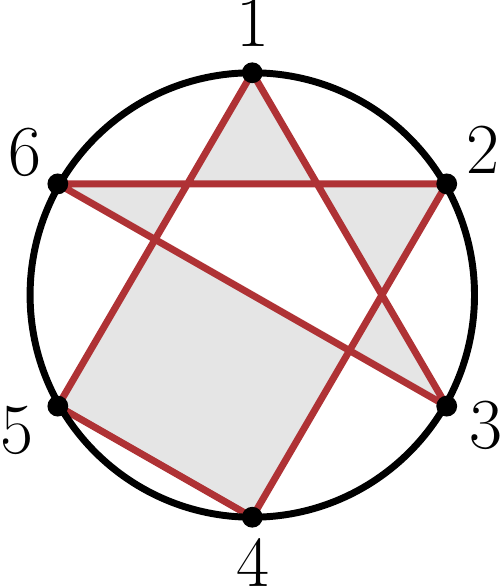}} & \parbox[c]{4em}{\includegraphics[scale=.3]{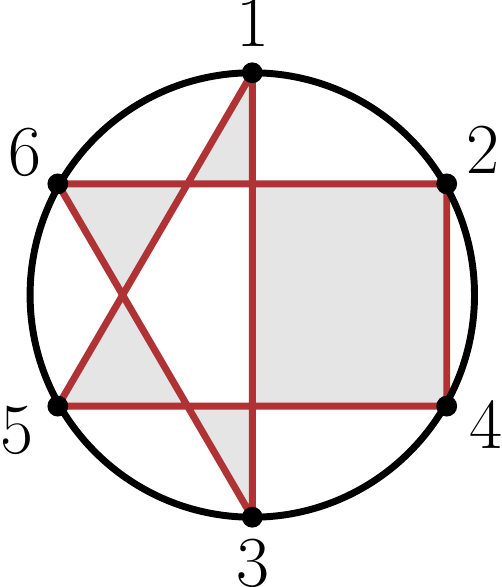}} & \parbox[c]{4em}{\includegraphics[scale=.3]{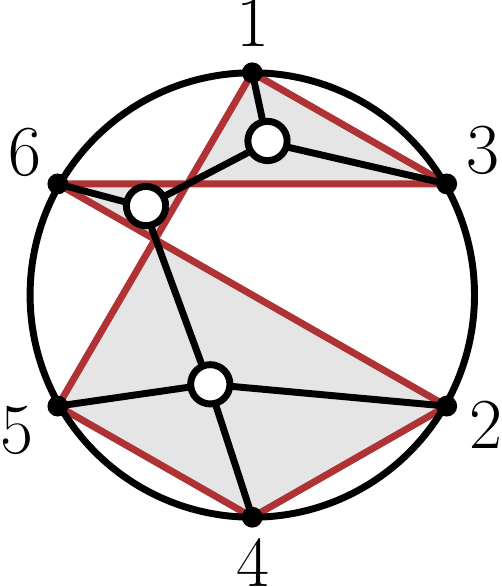}} & \parbox[c]{4em}{\includegraphics[scale=.3]{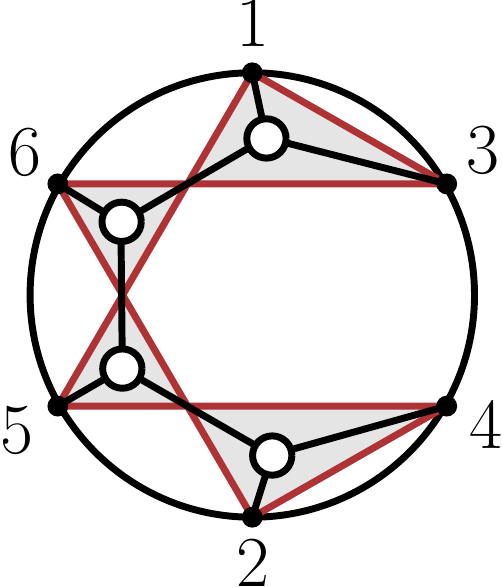}} & \parbox[c]{4em}{\includegraphics[scale=.3]{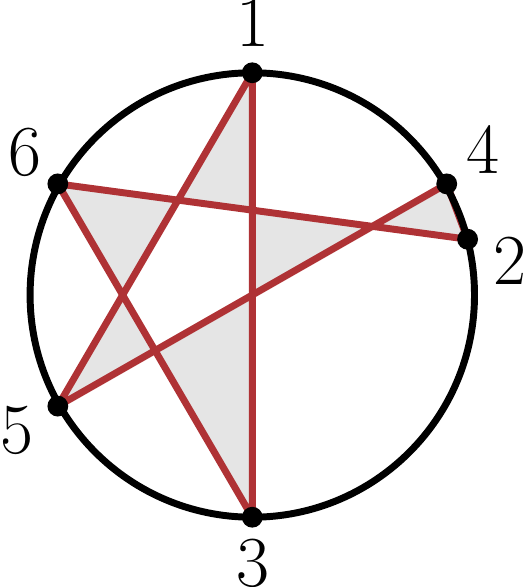}} &
\parbox[c]{4em}{\includegraphics[scale=.3]{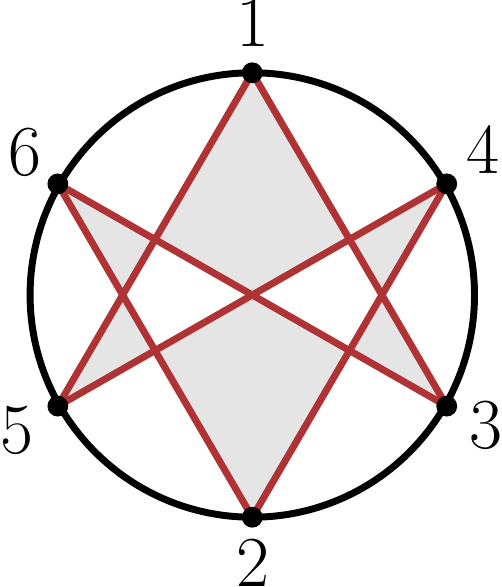}} \\ \vspace{-1em} \\
\parbox[c]{4em}{\includegraphics[scale=.3]{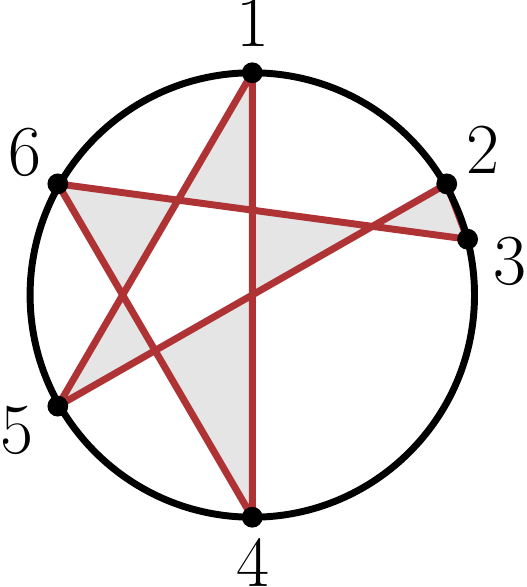}} & \parbox[c]{4em}{\includegraphics[scale=.3]{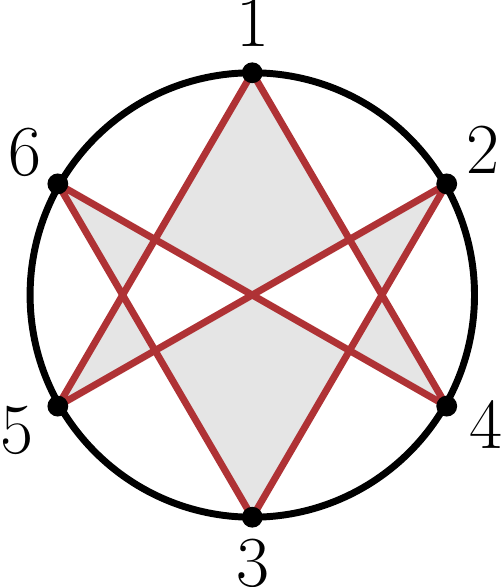}} & \parbox[c]{4em}{\includegraphics[scale=.3]{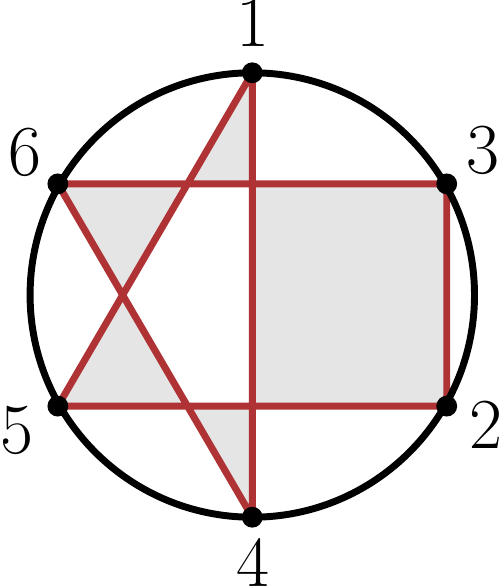}} & \parbox[c]{4em}{\includegraphics[scale=.3]{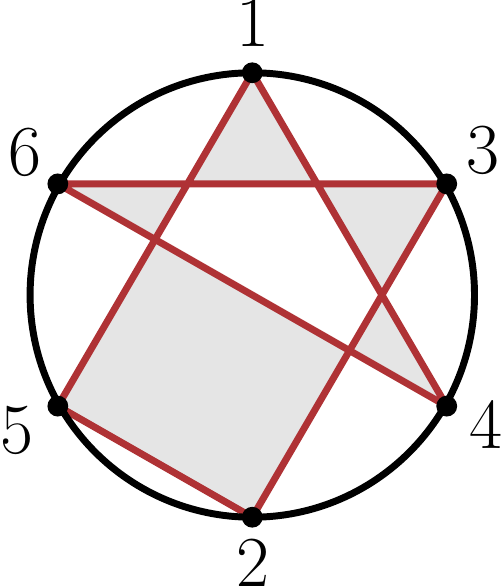}} & \parbox[c]{4em}{\includegraphics[scale=.3]{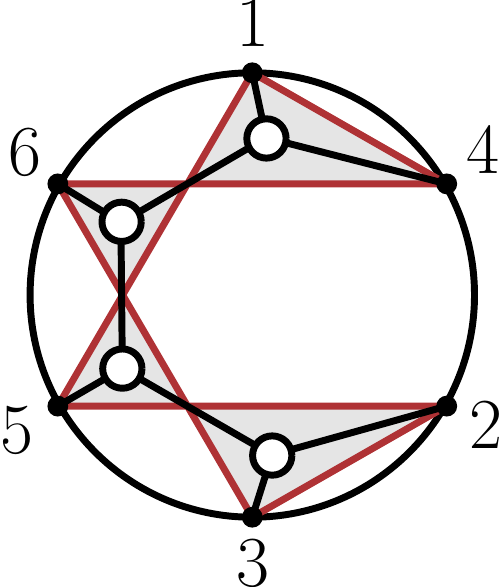}} &
\parbox[c]{4em}{\includegraphics[scale=.3]{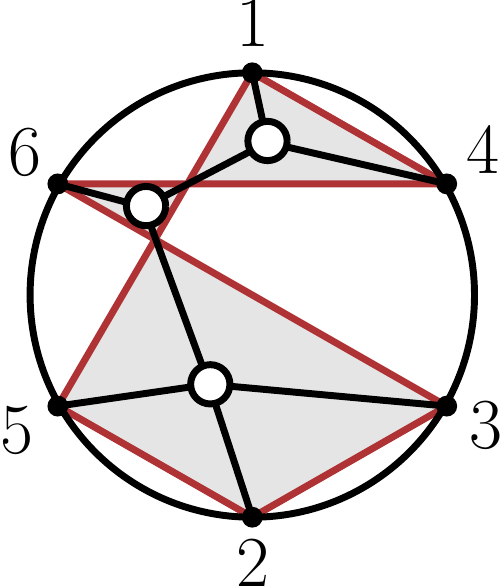}} \\ \vspace{-1em} \\
\parbox[c]{4em}{\includegraphics[scale=.3]{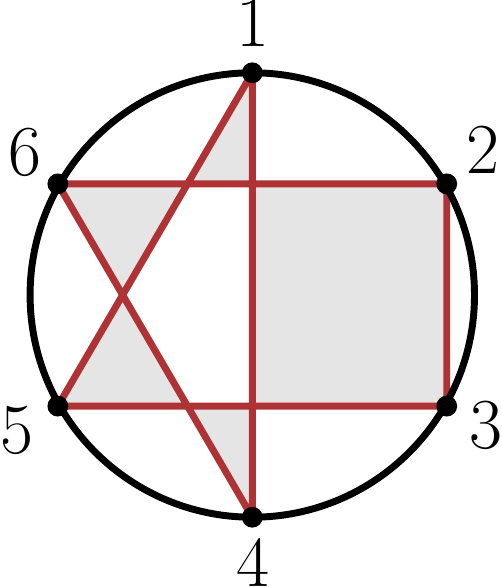}} & \parbox[c]{4em}{\includegraphics[scale=.3]{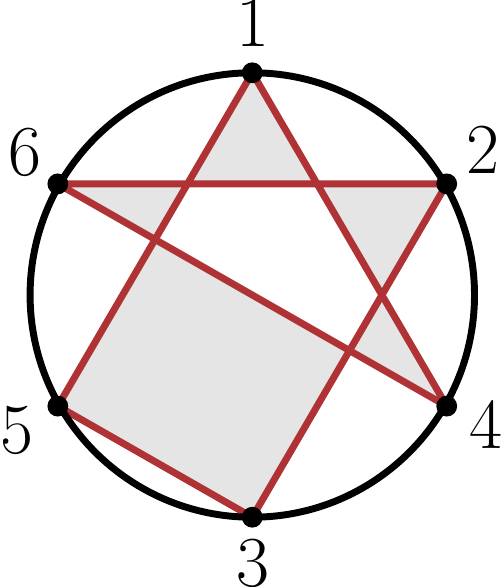}} & \parbox[c]{4em}{\includegraphics[scale=.3]{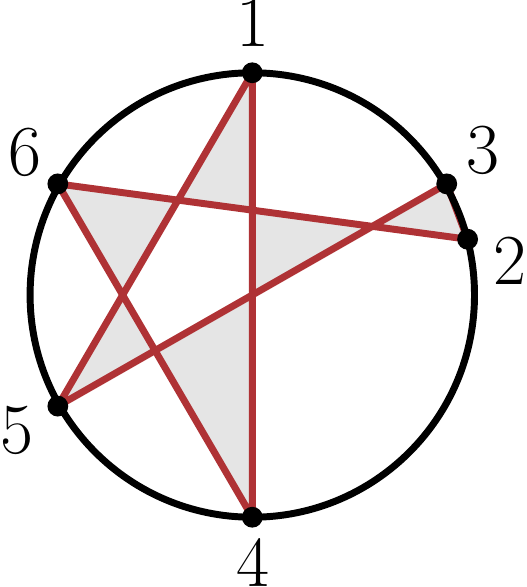}} & \parbox[c]{4em}{\includegraphics[scale=.3]{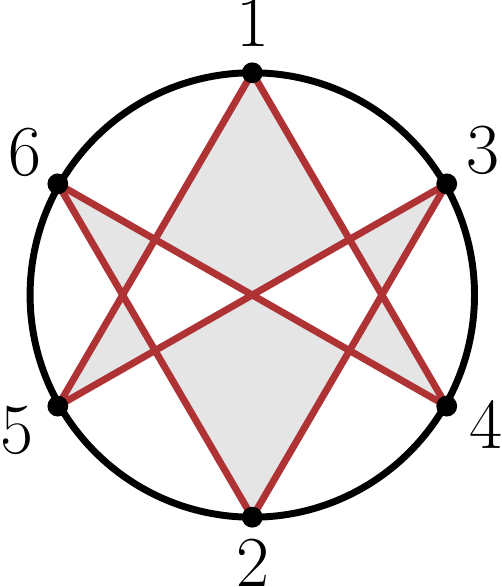}} & \parbox[c]{4em}{\includegraphics[scale=.3]{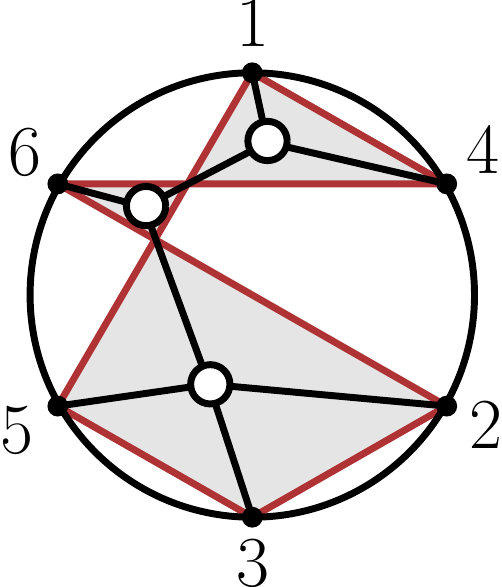}} &
\parbox[c]{4em}{\includegraphics[scale=.3]{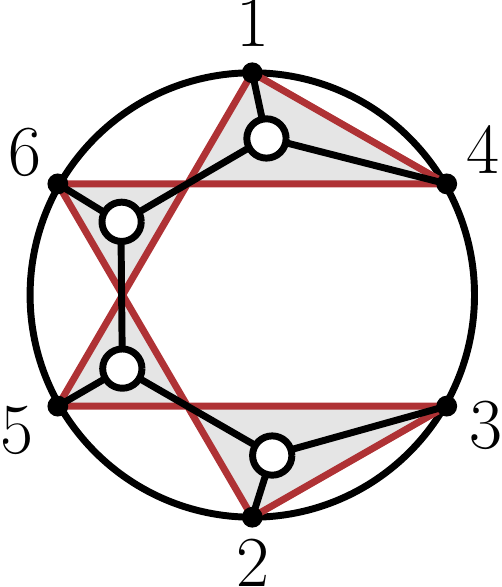}} \\ \vspace{-1em} \\
\end{bmatrix}.
\en
As one can see, components of this matrix are related by relabelling. It is then sufficient to calculate the inverse of the first block:

\bes
&&\begin{bmatrix}
\dfrac{1}{\sin s_{12}\, \sin s_{34}\, \sin s_{345}} & -\dfrac{1}{\sin s_{12}\, \sin s_{345}} \left(\dfrac{1}{\tan s_{34}} + \dfrac{1}{\tan s_{35}}\right) \\ \vspace{-1em} \\
-\dfrac{1}{\sin s_{12}\, \sin s_{345}} \left(\dfrac{1}{\tan s_{34}} + \dfrac{1}{\tan s_{45}}\right) & \dfrac{1}{\sin s_{12}\, \sin s_{34}\, \sin s_{345}}
\end{bmatrix}^{-1}\tr\tr
&&\qquad\qquad\qquad\!= \begin{bmatrix}
-\sin s_{12}\, \sin s_{35}\, \sin s_{45} & -\sin s_{12}\, \sin s_{45}\, \sin ( s_{34} + s_{35}) \\ \vspace{-1em} \\
-\sin s_{12}\, \sin s_{35}\, \sin ( s_{34} + s_{45}) & -\sin s_{12}\, \sin s_{35}\, \sin s_{45}
\end{bmatrix}.\quad
\ens
After contracting it with the two vectors of open amplitudes, we obtain the expression
\bes
\mathcal{M}^{\text{closed}}_6 &=& - \sin(\pi \alpha' s_{12})\, \sin (\pi \alpha' s_{45})\, \mathcal{A}^{\text{open}}(123456)\, \Big( \sin (\pi \alpha' s_{35})\, \mathcal{A}^{\text{open}}(153462) \tr
&& \qquad\qquad\qquad\qquad + \sin (\pi \alpha' (s_{34} + s_{35}))\, \mathcal{A}^{\text{open}}(154362) \Big) + \mathcal{P}(2,3,4),\qquad
\ens
which is the correct form of the KLT relation for $n=6$. Here, $\mathcal{P}(2,3,4)$ stands for the sum over permutations of $\{ 2,3,4\}$.

The rules for the computation of $m_{\alpha'}(\beta | \tilde{\beta})$ presented in this work, including the sign, can be automated. We have checked numerically that they reproduce entries in the KLT matrix in the form of \cite{BjerrumBohr:2010hn} up to $n=10$.

\subsection{\label{sec:Soft Limits}Soft Limits of Closed Strings}

As the first application of the newly-found interpretation of the KLT kernel, we compute the leading soft factor of the closed string amplitudes, given the knowledge of the open string soft factors. Closed string soft limits have been previously considered in various different ways \cite{Schwab:2014fia,Schwab:2014sla,DiVecchia:2015oba,DiVecchia:2016szw}. The leading soft factor is equal to the Weinberg soft factor for pure gravity \cite{Weinberg:1964ew, Weinberg:1965nx}. The novelty of our approach is that it does not require the precise knowledge of the KLT kernel, other than the diagrammatic rules of section \ref{sec:Diagrammatic Rules}.

We follow the construction of \cite{BjerrumBohr:2010hn,Cachazo:2016njl} by choosing the permutations labelled by two particles $a, b \neq 1,n-1,n$:
\be
\beta \in \{ (1, \omega_a, n-1, a, n) \} \quad \text{and} \quad \tilde{\beta} \in \{ (1, \tilde{\omega}_b, n-1, n, b) \},
\en
where $\omega_a$ and $\tilde{\omega}_b$ denote the permutations of the remaining $n-4$ labels. We can arrange the matrix $m_{\alpha'}(\beta | \tilde{\beta})$ so that $a$ and $b$ label its $(n-4)! \times (n-4)!$ blocks. Let us consider the behaviour of the $\alpha'$-corrected bi-adjoint amplitudes, with this specific ordering, under the soft limit of the particle $n$, $p_n = \tau \hat{p}_n$ with $\tau \to 0$. We keep $\alpha'$ finite.

In the case of the off-diagonal blocks, $a \neq b$, the particle $n$ is adjacent to different particles in both orderings $\beta$ and $\tilde{\beta}$. Therefore, there are no trivalent vertices involving the particle $n$ and hence all the propagators stay finite. We conclude that the off-diagonal blocks go as $\mathcal{O}(\tau^0)$ is the soft limit.

In the case of the diagonal blocks, $a = b$, the soft particle interacts with $a$ through a trivalent vertex, thus making the corresponding propagator diverge. Of course, it is also allowed to interact via higher-order vertices, but these give rise to finite contributions. Hence, the diagonal blocks behave as $\mathcal{O}(\tau^{-1})$ in the soft limit.

We have learned that the matrix $m_{\alpha'}(\beta | \tilde{\beta})$ becomes block diagonal in the soft limit. More precisely, near $\tau=0$ we have:
\be\label{eq:soft-biadjoint}
\hspace{-.5em}m_{\alpha'} (1, \omega_a, n-1, a, n| 1, \tilde{\omega}_b, n-1, n, b) \to -\frac{\delta_{ab}}{\tau \pi \alpha' \hat{s}_{an}} m_{\alpha'} (1, \omega_a, n-1, a | 1, \tilde{\omega}_a, n-1, a) + \ldots,
\en
where the minus sign comes about since the winding number changes by $1$. Each of the $(n-4)! \times (n-4)!$ diagonal blocks on the right hand side becomes a small inverse KLT matrix for $n-1$ particles.

We know that under the same soft limit, the open string partial amplitudes factorize as follows \cite{Mafra:2011nw}:
\be\label{eq:soft-open}
\mathcal{A}^{\text{open}} (1, \omega_a, n-1, a, n) \to \frac{1}{\tau}\left( \frac{\epsilon_n \cdot p_a}{\hat{p}_n \cdot p_a} - \frac{\epsilon_n \cdot p_1}{\hat{p}_n \cdot p_1} \right) \mathcal{A}^{\text{open}} (1, \omega_a, n-1, a) + \ldots,
\en
and similarly for the other ordering. We see that in each of the diagonal blocks, $a=b$, the bi-adjoint matrices \eqref{eq:soft-biadjoint} and the open string factors \eqref{eq:soft-open} combine to produce a closed string amplitude with $n-1$ particles. Working to leading order and neglecting constant factors, let us now collect all the terms together to obtain the result,
\bes
\mathcal{M}^{\text{closed}}_n &\to& -\frac{1}{\tau}\sum_{a=2}^{n-2} \left( \frac{\epsilon_n \cdot p_a}{\hat{p}_n \cdot p_a} - \frac{\epsilon_n \cdot p_1}{\hat{p}_n \cdot p_1} \right) \hat{s}_{an} \left( \frac{\tilde{\epsilon}_n \cdot p_{n-1}}{\hat{p}_n \cdot p_{n-1}} - \frac{\tilde{\epsilon}_n \cdot p_a}{\hat{p}_n \cdot p_a} \right) \mathcal{M}^{\text{closed}}_{n-1} + \ldots\tr
&=& \frac{1}{\tau} \left( \sum_{a=1}^{n-1} \frac{\epsilon_{\mu\nu}\, p_a^\mu\, p_a^\nu}{\hat{p}_n \cdot p_a} \right) \mathcal{M}^{\text{closed}}_{n-1}\, +\, \mathcal{O}(\tau^0),
\ens
where $\epsilon_{\mu\nu} = \epsilon_\mu \tilde{\epsilon}_\nu$ is the polarization vector of the massless state. In the last line we have used momentum conservation. This is indeed the correct soft factor for closed string amplitudes \cite{Schwab:2014fia}, coinciding with the pure gravity result.

\section{\label{sec:Basis Expansion}Basis Expansion}

Open string partial amplitudes can be expanded in a basis of size $(n-3)!$ \cite{BjerrumBohr:2009rd,Stieberger:2009hq}. This can be seen as a consequence of the monodromy relations \cite{BjerrumBohr:2009rd,Stieberger:2009hq,BjerrumBohr:2010zs}. In the infinite tension limit, the basis reduces to the BCJ relations \cite{Bern:2008qj} among Yang--Mills partial amplitudes. In this section, we show how to construct a basis expansion of open strings in an alternative way, utilizing the object $m_{\alpha'}(\beta | \tilde{\beta})$ introduced in this work.

Let us consider an $(n-3)!+1 \times (n-3)!+1$ matrix constructed from four blocks: an $(n-3)! \times (n-3)!$ matrix $m_{\alpha'}(\gamma|\tilde{\beta})$ with $\gamma$ and $\tilde{\beta}$ ranging in a set forming a BCJ basis, an $(n-3)!$ vector $\mathcal{A}^{\text{open}}(\gamma)$, and additionally $(n-3)!$ transposed vector $m_{\alpha'}(\beta | \tilde{\beta})$ and a scalar $\mathcal{A}^{\text{open}}(\beta)$ for a single permutation $\beta$. Since the first two objects form the same BCJ basis in $\gamma$, the latter two are linearly dependent. Hence the following determinant vanishes:
\be
\setlength{\arraycolsep}{3pt}
\det \left[ \begin{array}{c|c}
	m_{\alpha'}(\gamma | \tilde{\beta}) & \mathcal{A}^{\text{open}}(\gamma) \\
	\hline
	m_{\alpha'}(\beta | \tilde{\beta}) & \mathcal{A}^{\text{open}}(\beta) \\
\end{array} \right] = \left( \mathcal{A}^{\text{open}}(\beta) - m_{\alpha'}(\beta | \tilde{\beta})\, m_{\alpha'}^{-1}(\tilde{\beta} | \gamma)\, \mathcal{A}^{\text{open}}(\gamma) \right) \det m_{\alpha'}(\gamma | \tilde{\beta}) = 0,\nn
\en
where we have used a determinant expansion and Sylvester's determinant theorem \cite{MatrixManual} in the first equality. Since the sets over which $\beta$ and $\tilde{\beta}$ range form a BCJ basis, the determinant of $m_{\alpha'}(\gamma | \tilde{\beta})$ is non-vanishing, so we conclude that:
\bes\label{eq:open-basis}
\mathcal{A}^{\text{open}}(\beta) = m^{}_{\alpha'}(\beta | \tilde{\beta})\, m^{-1}_{\alpha'}(\tilde{\beta} | \gamma)\, \mathcal{A}^{\text{open}}(\gamma),
\ens
which is a basis expansion for the open string amplitude $\mathcal{A}^{\text{open}}(\beta)$ in terms of a BCJ basis. It is a direct analogue of the expansion \eqref{eq:basis-expansion} for the field-theory amplitudes. Using the same argument one can prove similar relations for other BCJ-satisfying amplitudes. In the following, we illustrate the expansion \eqref{eq:open-basis} with a few examples.

As a first example for $n=4$, we expand $\mathcal{A}^{\text{open}}(1243)$ in the one-element basis $\{ \mathcal{A}^{\text{open} }(1234) \}$. This corresponds to taking $\beta = \{(1243)\}$, $\gamma = \{(1234)\}$, and say $\tilde{\beta} = \{(1324)\}$. We obtain:
\bes\label{eq:Aopen-1243}
\mathcal{A}^{\text{open}} (1243) &=& \left(\; \parbox[c]{5.1em}{\includegraphics[scale=.4]{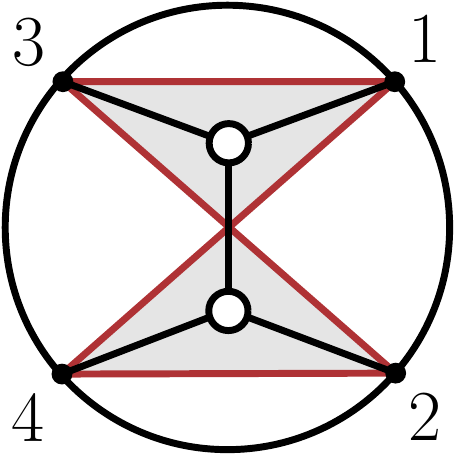}} \right) \left(\; \parbox[c]{5.1em}{\includegraphics[scale=.4]{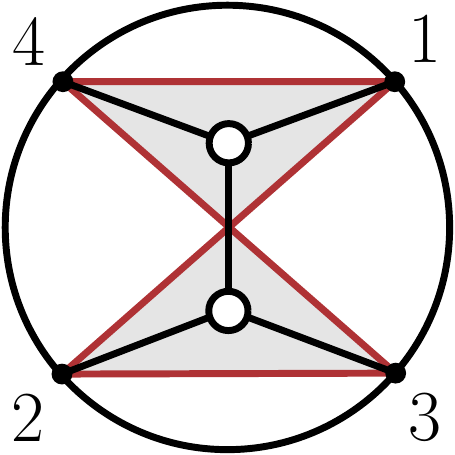}} \right)^{-1}\!\!\!\! \mathcal{A}^{\text{open}}(1234)\tr\tr
&=& \frac{\sin (\pi \alpha' t)}{\sin (\pi \alpha' u)}\, \mathcal{A}^{\text{open}}(1234).
\ens
A slightly more involved example is:
\bes\label{eq:Aopen-1324}
\mathcal{A}^{\text{open}} (1324) &=& \left(\; \parbox[c]{5.1em}{\includegraphics[scale=.4]{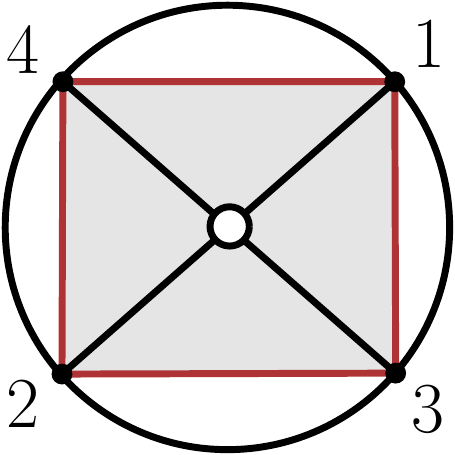}} \right) \left(\; \parbox[c]{5.1em}{\includegraphics[scale=.4]{figures/m-1324-1234}} \right)^{-1}\!\!\!\! \mathcal{A}^{\text{open}}(1234)\tr\tr
&=& \left( \frac{1}{\tan (\pi \alpha' u)} + \frac{1}{\tan (\pi \alpha' t)} \right) \left( -\sin (\pi \alpha' t) \right)\, \mathcal{A}^{\text{open}}(1234)\tr
&=& \frac{\sin (\pi \alpha' s)}{\sin (\pi \alpha' u)}\, \mathcal{A}^{\text{open}}(1234).
\ens
These two examples can be easily verified using the monodromy relations \cite{BjerrumBohr:2009rd}. Both cases yield the correct field theory limit. Of course, in practical calculations we would put both of the above amplitudes \eqref{eq:Aopen-1243} and \eqref{eq:Aopen-1324} into a vector, so that the KLT kernel has to be inverted only once.

As the closing example, let us generalize the five point Yang--Mills basis expansion \eqref{eq:basis5} to string theory. Making the choice $\beta = \{(12354)\}$, $\tilde{\beta} = \{ (12345), (12435) \}$ and $\gamma = \{ (13254), (14253) \}$ we get:
\bes
&&\mathcal{A}^{\text{open}}(12354) = \begin{bmatrix} \;\,\parbox[c]{5.8em}{\includegraphics[scale=.4]{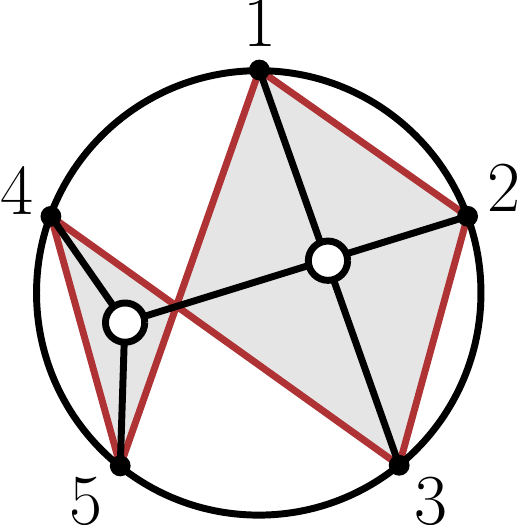}} \\ \;\,\parbox[c]{5.8em}{\includegraphics[scale=.4]{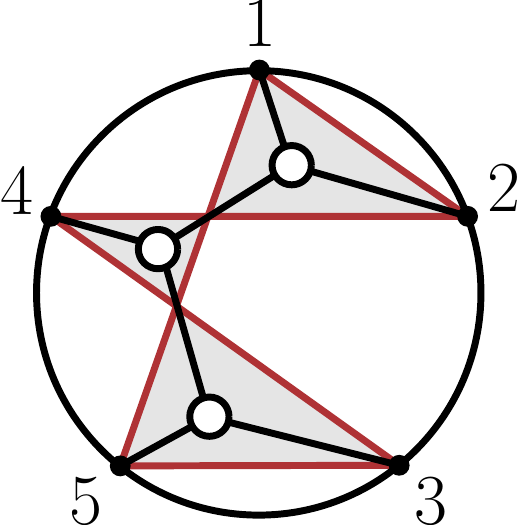}} \end{bmatrix}^\intercal \begin{bmatrix} \;\,\parbox[c]{5.8em}{\includegraphics[scale=.4]{figures/m-12345-13254}} & \parbox[c]{5.8em}{\includegraphics[scale=.4]{figures/m-12435-14253}} \\ \vspace{-.9em}\\ \;\,\parbox[c]{5.8em}{\includegraphics[scale=.4]{figures/m-12345-14253}} & \parbox[c]{5.8em}{\includegraphics[scale=.4]{figures/m-12435-13254}} \end{bmatrix}^{-1} \begin{bmatrix} \mathcal{A}^{\text{open}}(13254)\\ \mathcal{A}^{\text{open}}(14253) \end{bmatrix}\tr
&&\!\!\!= \begin{bmatrix} -\dfrac{1}{\sin s_{45}} \left( \dfrac{1}{\tan s_{12}} + \dfrac{1}{\tan s_{23}} \right) \\ - \dfrac{1}{\sin s_{12} \sin s_{35}} \end{bmatrix}^\intercal \begin{bmatrix} \dfrac{1}{\sin s_{23} \, \sin s_{45}} & 0 \\ 0 & \dfrac{1}{\sin s_{24}\, \sin s_{35}} \end{bmatrix}^{-1} \begin{bmatrix} \mathcal{A}^{\text{open}}(13254)\\ \mathcal{A}^{\text{open}}(14253) \end{bmatrix}\tr\tr
&&\!\!\!= - \frac{\sin (\pi \alpha' (s_{12} + s_{23}))}{\sin (\pi \alpha' s_{12})} \,\mathcal{A}^{\text{open}}(13254) - \frac{\sin (\pi \alpha' s_{24})}{\sin (\pi \alpha' s_{12})} \,\mathcal{A}^{\text{open}}(14253),
\ens
which can be verified against the explicit form of the open string amplitudes, e.g., \cite{Medina:2002nk}.

\section{\label{sec:Future Directions}Future Directions}

In this work, we have shown that the inverse of the string theory KLT kernel can be understood as a matrix of amplitudes in an $\alpha'$-corrected bi-adjoint theory, $m_{\alpha'}(\beta | \tilde{\beta})$. Given that a closed string amplitude is calculated from a correlation function of vertex operators on a genus-$0$ Riemann surface, and an open string amplitude comes from a correlator of operators inserted on a disk boundary of a genus-$0$ Riemann surface, the KLT relations \eqref{eq:string KLT} can be graphically summarized in a cartoon:
\vspace{0.5em}
\begin{center}
\includegraphics[scale=.75]{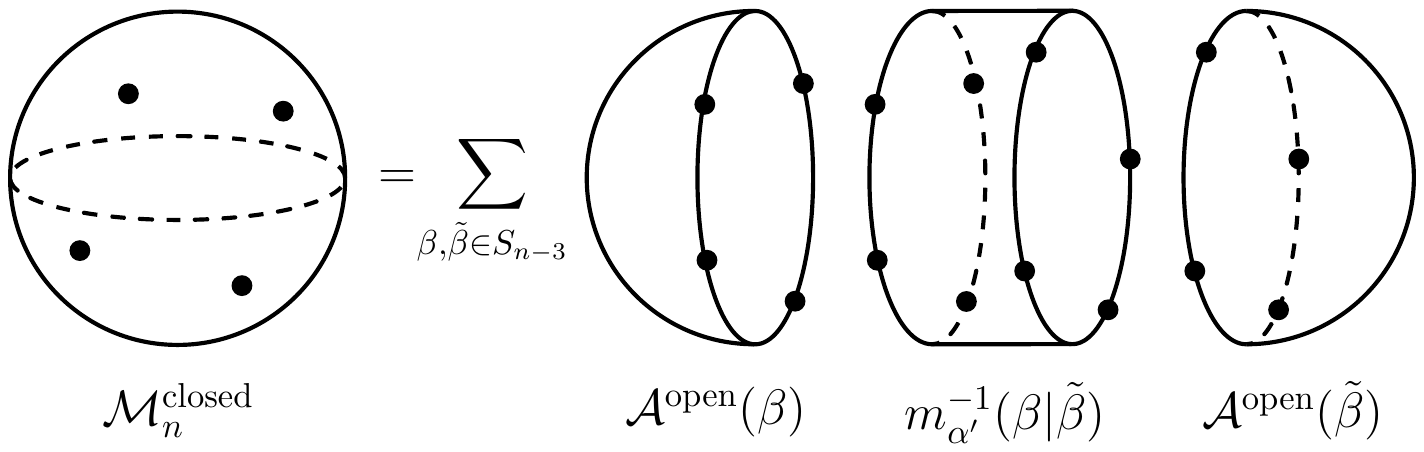}
\end{center}
Here, the closed string amplitude $\mathcal{M}_{n}^{\mathrm{closed}}$ is understood as gluing of two open string partial amplitudes, $\mathcal{A}^{\mathrm{open}}(\beta)$ and $\mathcal{A}^{\mathrm{open}}(\tilde{\beta})$. The object stitching the two amplitudes together is the KLT kernel, $m_{\alpha'}^{-1}(\beta | \tilde{\beta})$. The sum proceeds over all independent ways of performing the gluing, with $\beta$ and $\tilde{\beta}$ each ranging over a set of $(n-3)!$ permutations. Similar picture can be made to intuitively understand the change of basis relation \eqref{eq:open-basis}.

This interpretation of the KLT kernel departs from the original understanding \cite{Kawai:1985xq} of the coefficients of KLT expansion as simply coming from a contour deformation argument, where the factors of sines arise from monodromy properties of string integrals. We have argued that $m_{\alpha'}(\beta | \tilde{\beta})$ can be understood as an interesting object on its own right. In fact, we propose that it could itself originate from a string integral with two disk orderings. We present some supporting evidence for this conjecture below.

The main results of this work concerns simplicity of the inverse string theory KLT kernel. We have shown how the objects $m_{\alpha'}(\beta | \tilde{\beta})$ take a compact form, which is amenable to a diagrammatic expansion. This fact has an immediate application in expanding open string amplitudes in BCJ bases, as described in section \ref{sec:Basis Expansion}. As it is usual in the study of scattering amplitudes, such simplicity hints at the existence of some underlying structure. Perhaps it can be related to the work on planar algebras \cite{Ceyhan:2009xg} or intersection matrices associated to Selberg integrals \cite{Mimachi2003,mimachi2004}.\footnote{We thank Matilde Marcolli and Oliver Schlotterer for pointing out these references to us.}

Additionally, the field theory bi-adjoint amplitudes have a CHY representation that allowed for its identification with the inverse KLT kernel in the first place \cite{Cachazo:2013iea}. In an ongoing program of finding connections between the CHY formalism and string theory \cite{Mason:2013sva,Adamo:2013tsa,Gomez:2013wza,Geyer:2014fka,Bjerrum-Bohr:2014qwa,Casali:2015vta,Casali:2016atr,He:2016iqi,Cachazo:2016ror}, understanding of the $\alpha'$-corrected KLT procedure in the CHY language could shed a new light on the relations between the two.

We can relate the $\alpha'$-corrected bi-adjoint theory to other objects studied in the context of string amplitudes. Our starting point is an intriguing new identity conjectured by Huang, Siegel and Yuan in \cite{Huang:2016bdd}. The claim is that supergravity amplitudes can be obtained from a string theory KLT of open string amplitudes with a small ``twist,''
\bes\label{eq:SUGRA-open}
\mathcal{M}^{\text{SUGRA}}_n = \mathcal{A}^{\text{open}} \KLTs\, \overline{\mathcal{A}^{\text{open}}}.
\ens
Here, the bar over the second open string vector denotes a flip of the space-time signature, also equivalent to a change $\alpha' \to -\alpha'$. In addition, it is known \cite{Mafra:2011nv,Mafra:2011nw,Broedel:2013aza,Broedel:2013tta} that open string amplitudes can be expanded in a basis of the super Yang--Mills ones as follows:
\bes\label{eq:open-Z-SYM}
\mathcal{A}^{\text{open}}(\beta) = Z_{\beta} \KLT \mathcal{A}^{\text{SYM}},
\ens
where $Z_\beta (\gamma)$ is an amplitude of a string-like theory \cite{Carrasco:2016ldy,Mafra:2016mcc} carrying all the $\alpha'$ dependence. Its disk integral representation, up to an overall factor, reads:
\bes\label{eq:Z-defn}
Z_\beta (\gamma) \;= \!\!\!\!\!\!\!\!\!\! \int\displaylimits_{z_{\beta_1} < z_{\beta_2} < \cdots < z_{\beta_n}} \!\!\!\!\!\!\!\!\!\! \frac{d^n z}{\text{vol SL}(2,\mathbb{R})} \, \frac{\prod_{i<j} |z_{ij}|^{\alpha' s_{ij}}}{z_{\gamma_1,\gamma_2} \, z_{\gamma_2,\gamma_3}\, \cdots\, z_{\gamma_{n-1},\gamma_n}\, z_{\gamma_{n},\gamma_1}}.
\ens
Note that the two orderings play different roles. The permutation $\beta$ gives a disk ordering that is inherited by the open string. The ordering $\gamma$ enters the Parke-Taylor factor in the integrand and eventually gets contracted in the KLT relation \eqref{eq:open-Z-SYM}. Recalling that supergravity and super Yang--Mills are related by the usual field theory KLT relation, $\mathcal{M}^{\text{SUGRA}} = \mathcal{A}^{\text{SYM}} \KLT \mathcal{A}^{\text{SYM}}$, it is a simple linear algebra exercise to combine it with \eqref{eq:SUGRA-open} and \eqref{eq:open-Z-SYM} to obtain:
\bes\label{eq:m-alpha-Z}
m(\gamma | \tilde{\gamma}) = Z(\gamma) \KLTs \overline{Z(\tilde{\gamma})} \qquad\text{or equivalently}\qquad m_{\alpha'}(\beta | \tilde{\beta}) = Z^{}_{\beta} \KLT \overline{Z_{\tilde{\beta}}}.
\ens
Here the overbar has the same meaning as in \eqref{eq:SUGRA-open}. It is also possible to define the $\alpha'$-corrected bi-adjoint theory in terms of other string-like objects defined in \cite{Stieberger:2014hba,Huang:2016tag}.

Due to the relation \eqref{eq:m-alpha-Z}, the $\alpha'$-corrected bi-adjoint amplitudes $m_{\alpha'}(\beta | \tilde{\beta})$ inherit some symmetries of the integrals $Z_{\beta}(\gamma)$, namely, the cyclicity and parity properties in the orderings $\beta$ and $\tilde{\beta}$, as well as monodromy relations \cite{Mafra:2011nw} in both orderings separately. Since monodromy relations are a consequence of the disk integrals, this fact provides support to the claim that $m_{\alpha'}(\beta | \tilde{\beta})$ should have a representation as a string integral with two disk orderings.

Furthermore, using \eqref{eq:open-Z-SYM}, one can show that \eqref{eq:open-basis} implies,
\be
Z_{\beta}(\gamma) = m^{}_{\alpha'}(\beta | \tilde{\beta})\, m^{-1}_{\alpha'}(\tilde{\beta} | \delta)\, Z_{\delta} (\gamma).
\en
That is, the new object provides a way of changing a basis of the disk ordering in $Z_{\beta}(\gamma)$, which is distinct from the change of basis for the other permutation \cite{Mafra:2011nw},
\be
Z_{\beta}(\gamma) = m(\gamma | \tilde{\gamma})\, m^{-1}(\tilde{\gamma} | \varepsilon)\, Z_{\beta} (\varepsilon).
\en

The connection between the field theory and string KLT kernels has already been studied from the perspective of multiple zeta values \cite{Stieberger:2009rr,Mafra:2011nv,Mafra:2011nw,Broedel:2013aza,Broedel:2013tta,Schlotterer:2012ny,Stieberger:2013wea,Stieberger:2014hba,Huang:2016tag}. Inverting the result from \cite{Schlotterer:2012ny}, one finds,
\be
m_{\alpha'}(\beta | \tilde{\beta}) = P(\beta | \gamma)\, m(\gamma | \tilde{\gamma})\, P(\tilde{\gamma} | \tilde{\beta}),
\en
where
\be\label{eq:P}
P(\beta | \gamma) = \delta_{\beta \gamma} + \sum_{n \geq 1} \zeta_2^n\; Z_{\beta}(\delta) \bigg|_{\zeta_2^n} \!\! m^{-1}(\delta | \gamma).
\en
The connection to \eqref{eq:SUGRA-open} has been examined in \cite{Huang:2016bdd}. An interesting question is how to obtain the compact expressions for $m_{\alpha'}(\beta | \tilde{\beta})$ presented in this work from the motivic structure of \eqref{eq:P}.

Finally, in this work we have not pursued questions pertaining to the colour-kinemetics duality. The field theory bi-adjoint scalar plays a prominent role in understanding of this duality both on-shell \cite{BjerrumBohr:2012mg,Cachazo:2013gna,Cachazo:2013iea,Monteiro:2013rya,Chiodaroli:2015rdg,Bjerrum-Bohr:2016axv} and off-shell \cite{Monteiro:2014cda,Luna:2015paa,Luna:2016due,Ridgway:2015fdl}. It would be very interesting to study the $\alpha'$-corrected version of these developments in the light of the results presented in this note.

\section*{Acknowledgements}

It is a pleasure to thank Freddy Cachazo for a careful reading of this manuscript and helpful comments. We also thank C. Kalousios, M. Marcolli, O. Schlotterer and G. Zhang for useful discussions. This research was supported in part by Perimeter Institute for Theoretical Physics. Research at Perimeter Institute is supported by the Government of Canada through the Department of Innovation, Science and Economic Development Canada and by the Province of Ontario through the Ministry of Research, Innovation and Science.

\begin{appendices}

\section{\label{sec:Proof of the Sign}Proof of the Sign of $m_{\alpha'}(\beta|\tilde{\beta})$}

In section \ref{sec:Off-diagonal amplitudes}, we gave a prescription for computing the sign of $m_{\alpha'}(\beta | \tilde{\beta})$. It says that
\be\label{eq:sign-definition}
\text{sgn} ( m_{\alpha'}(\beta | \tilde{\beta}) ) = (-1)^{w(\beta | \tilde{\beta}) + 1},
\en
where $w(\beta | \tilde{\beta})$ is the relative winding number between permutations $\beta$ and $\tilde{\beta}$, see \eqref{eq:winding} for an example. Since the overall sign of $m_{\alpha'}(\beta | \tilde{\beta})$ does not change in the $\alpha' \to 0$ limit, it is sufficient to show that \eqref{eq:sign-definition} works for the field theory bi-adjoint scalar.

Let us consider any Feynman diagram consistent with both orderings, $\beta$ and $\tilde{\beta}$. Each vertex carries a factor $f^{abc} \tilde{f}^{\tilde{a} \tilde{b} \tilde{c}}$. There are two options. When the labels of the three legs belonging to the vertex are the same in both orderings modulo cyclicity, the vertex contributes a plus sign to the trace decomposition \eqref{eq:Trace decomposition}. Otherwise, it contributes a minus sign.

We can now reshuffle the permutation $\tilde{\beta}$ into $\beta$ by a series of \emph{flips}. A single flip is a change of the labels of $\tilde{\beta}$, so that one vertex changes sign, e.g. from $(abc|acb)$ to $(abc|abc)$. In a finite number of steps, we reach a configuration with all plus signs on the vertices, which necessarily corresponds to $\beta = \tilde{\beta}$. Since this configuration comes with the plus sign, the overall sign of the initial configuration is $(-1)^{\text{\#flips}}$.

Let us analyze the situation from the perspective of the winding number definition \eqref{eq:sign-definition}. For the diagonal configuration, $\beta=\tilde{\beta}$, the winding number is one and hence the sign is a plus. Each of the flips changes the winding number by exactly one. Therefore, the definition \eqref{eq:sign-definition} computes the correct sign.
	
\end{appendices}

\bibliographystyle{JHEP}
\bibliography{references}

\end{document}